\documentclass[12pt,letterpaper]{article}

\usepackage{graphicx}
\usepackage{epstopdf}
\usepackage[body={17.5cm, 22cm},right=2.2cm]{geometry}
\usepackage{amssymb}
\usepackage{amsmath} 

\newcommand\ee{\end{equation}}
\newcommand\be{\begin{equation}}
\newcommand\eea{\end{eqnarray}}
\newcommand\bea{\begin{eqnarray}}
\newcommand{\sfrac}[2]{{\textstyle\frac{#1}{#2}}}
\newcommand\di{\partial}
\newcommand\mpl{M_{\rm Pl}}
\newcommand{\newc}{\newcommand}
\newc{\gsim}{\lower.7ex\hbox{$\;\stackrel{\textstyle>}{\sim}\;$}}
\newc{\lsim}{\lower.7ex\hbox{$\;\stackrel{\textstyle<}{\sim}\;$}}

\begin{document}

\setcounter{page}{0}
\thispagestyle{empty}

\begin{titlepage}

\begin{center}

\vspace{1.cm}

{\LARGE \sc{
The galileon as a local \\[.4cm] modification of gravity
}}\\[1.5cm]
{\large Alberto Nicolis$^{\rm a}$, Riccardo Rattazzi$^{\rm b}$, Enrico Trincherini$^{\rm c}$}
\\[.8cm]

\vspace{.2cm}
{\small \textit{$^{\rm a}$ 
Department of Physics and ISCAP,\\
 Columbia University, 
New York, NY 10027, USA }}

\vspace{.2cm}
{\small \textit{$^{\rm b}$ 
Institut de Th\'eorie des Ph\'enom\`enes Physiques, EPFL,\\
CHÐ1015 Lausanne, Switzerland}}        

\vspace{.2cm}
{\small \textit{$^{\rm c}$ 
Scuola Normale Superiore, \\
Piazza dei Cavalieri 7, 56126 Pisa, Italy}}

\end{center}

\vspace{.8cm}
\begin{abstract}

In the DGP model, the ``self-accelerating'' solution is plagued by a  
ghost instability, which makes the solution untenable.
This fact as well as all interesting departures from GR are fully  
captured by a four-dimensional effective Lagrangian, valid at  
distances smaller than the present Hubble scale. The 4D effective  
theory involves a relativistic scalar $\pi$, universally coupled to  
matter and with peculiar derivative self-interactions.
In this paper, we study the connection between self-acceleration and  
the presence of ghosts for a quite generic class of theories that  
modify gravity in the infrared. These theories are defined as those  
that at distances shorter than cosmological, reduce to a certain  
generalization of the DGP 4D effective theory. We argue that for  
infrared modifications of GR locally due to a universally coupled  
scalar, our generalization is the only one that allows for a robust  
implementation of the Vainshtein effect---the decoupling of the  
scalar from matter in gravitationally bound systems---necessary to  
recover agreement with solar system tests. Our generalization  
involves an internal  ``galilean'' invariance, under which $\pi$'s  
{\em gradient} shifts by a constant. This symmetry constrains the  
structure of the $\pi$ Lagrangian so much so that in 4D there exist  
only five terms that can yield sizable non-linearities without  
introducing ghosts.
We show that for such theories in fact there are ``self-accelerating'' deSitter solutions with no ghost-like instabilities.  
In the presence of compact sources, these solutions can support  
spherically symmetric, Vainshtein-like non-linear perturbations that  
are also stable against small fluctuations. We investigate a possible {\em infrared} completion of these theories at scales of order of the Hubble horizon, and larger.
There are however some features of our theories, that may constitute  
a problem at the theoretical or phenomenological level: the presence  
of superluminal excitations;  the extreme {\em sub}-luminality of  
other excitations, which makes the quasi-static approximation for  
certain solar-system observables unreliable due to Cherenkov  
emission; the very low strong-interaction scale for $\pi \pi$  
scatterings.

\end{abstract}

\end{titlepage}


\section*{Introduction}

The beauty of General Relativity (GR) lies in the simplicity of its principles and in its  empirical adequacy in a broad range
of phenomena. The modern viewpoint is that GR is an effective theory valid below some ultraviolet cut-off, presumably of order $M_P\sim 10^{19}$ GeV,  and that its validity becomes increasingly better the lower the energy, that is the longer the distance scales. In particular this viewpoint explains the adequacy of GR at macroscopic and solar system scales and implies
also its perfect adequacy to describe cosmology at late enough times (and at present). However the undeniable beauty of GR has been stained, from almost the beginning, by the mystery of the cosmological constant. The inability to provide, to date, a plausible field theory mechanism for the absence or smallness of the cosmological constant, perhaps indicates that something is missing in the simple infrared picture of GR. Indeed, the strong evidence accumulated in recent years that the universe is  undergoing a phase of accelerated expansion  \cite{Riess:1998cb} provides extra motivation to consider a modification of gravitational dynamics
on cosmological scales. Could late time acceleration follow from the very dynamical mechanism that screens the cosmological constant? At the moment we have no concrete example in that sense. On the other hand there is instead a plausible, and even remarkable, picture where GR is the right infrared decription and where the size of the cosmological constant is just explained by anthropic selection \cite{Weinberg:1987dv}. Nonetheless the big effort
of recent years in trying to come up with, and to understand, modifications of gravity has produced an interesting know-how.
We have now several examples of theories that modify gravity \cite{ArkaniHamed:2003uy,Rubakov:2004eb,Dubovsky:2004sg,Dvali:2000hr}. In a broad technical sense we may define a modification of gravity
as a field theory possessing solutions over which new degrees of freedom affect the propagation of gravity while the background is not producing any sizable energy momentum tensor \cite{Dubovsky:2005xd}. In other words in these theories the graviton propagates differently even in flat space.
Examples of this type based on purely 4-dimensional field theory are the theories of massive gravity, like Fierz-Pauli or more potentially interesting models based on the mechanism of ghost condensation \cite{ArkaniHamed:2003uy}.
 A perhaps more remarkable example is however represented by the Dvali-Gabadadze-Porrati (DGP) model \cite{Dvali:2000hr}, based on 5-dimensional
 spacetime. The DGP model is interesting for several reasons. It cannot truly be assimilated to a fluid, as the infrared dynamics
involves  a continuum of states, corresponding to space-time being 5-dimensional, not just a few phonons. It potentially represents  a stepping stone towards the realization of the old idea of diluting the cosmological constant into extra dimensions \cite{Rubakov:1983bz,Dvali:2002pe}.
 It provides an alternative explanation for the observed late time acceleration of the universe \cite{deffayet}.  Indeed it nicely illustrates that in order to curve space-time at long distances  it is not necessary to add a zero derivative term, that is the cosmological constant, while it is `enough' a one-derivative term. The latter is in DGP formally represented by the 5 dimensional kinetic term, which is more `relevant' in the tecnical sense than the 4-D Einstein term localized at the boundary. Yet another reason of interest in the DGP model lies
 in its successful implementation of the Vainshtein mechanism \cite{Vainshtein:1972sx} for decoupling the additional modes from  gravitational dynamics at sub-cosmological distances. The problem at hand, which was first pointed out by van Dam, Veltman and Zacharov \cite{vanDam:1970vg}, is tied to the fact that there is no such thing as a purely IR modification of gravity: degrees of freedom with
 mass of order of the Hubble parameter, can expectedly play a role at shorter distances. Indeed, in both Fierz-Pauli and DGP
 these short distance effects are accurately described just by one additional scalar degree of freedom \cite{ArkaniHamed:2002sp,LPR}, normally indicated by $\pi$. The $\pi$ modifies at $O(1)$ the linearized gravitational potential produced by any energy momentum source. Tests of GR within the solar system rule out such $O(1)$ effects, and seemingly the models that produce them. As first remarked by Vainshtein long ago,
 non-linear effects in the dynamics of $\pi$  are however important. Now, it turns out that in the case of Fierz Pauli
 this does not lead to any benefits, as the non linearities always imply the presence of a light ghost state and the consequent
 breakdown of  effective field theory \cite{Creminelli:2005qk,Deffayet:2005ys}. However in the case of DGP, under an intersting range of conditions on the energy momentum tensor, the Vainshtein effect screens the contribution of $\pi$  below the experimentally acceptable level \cite{Deffayet:2001uk, Gruzinov:2001hp, Porrati:2002cp} without introducing extra ghost states. Unfortunately the interesting self-accelerating solution is not in this class, in that the $\pi$ is a ghost already on unperturbed de Sitter \cite{LPR, NR,Gorbunov:2005zk}.  Now, one crucial property of the $\pi$ Lagrangian ${\cal L}_\pi$ in DGP, the one that gives the Vainshtein effect a chance to work,  is that, while involving higher derivative interactions, it nonetheless gives equations of motion of second order in the derivatives \cite{NR}. Indeed all of DGP physics at distances shorter than $H^{-1}$, including the existence of the self-accelerating solution,
is correctly reproduced by just the $\pi$ field coupled to 4D gravity, without any reference to its 5D origin. These facts motivate
a classification and study of all the purely 4D lagrangians involving just one scalar field $\pi$ coupled to gravity and to itself in the same peculiar way as in DGP. In particular this approach will be enough to study the possible correlation between self-acceleration and existence of ghosts, or else to find self-accelerating solutions that lack ghosts. This will be the main goal of this paper.


\section{Local analysis}

\subsection{Cosmology as a weak gravitational field \label{cosmology}}

A FRW universe is a highly non-linear solution of Einstein's equations.
Still, at short scales---shorter than the Hubble radius---a cosmological solution can certainly be thought of as a small perturbation of Minkowski space.
For instance, given a comoving observer sitting at the origin, we can choose coordinates that for all times make any deviation from Minkowski metric weighted by $\sim H^2 |\vec x|^2$, where $|\vec x|$ is the physical distance away from the observer. 
Such a coordinate choice is not unique. A particularly convenient one is Newtonian gauge, for which $g_{0i}$ vanishes and the spatial metric is isotropic,
\be \label{Newtonian}
ds^2 = - \big( 1 + 2 \Phi \big) dt^2 + \big( 1 - 2 \Psi \big) d \vec x ^2 \; .
\ee
More explicitly, if $\tau$ and $\vec y$ are FRW comoving coordinates,
\be
ds^2 = -d \tau^2 + a^2(\tau) d \vec y ^2  \; ,
\ee
we perform the coordinate change
\be
\tau = t  - \sfrac{1}{2} H |\vec x|^2  \; , \qquad \vec y = \frac{\vec x}{a} \big[ 1 + \sfrac{1}{4} H^2 |\vec x|^2 \big] \; ,
\ee
where $a$ and $H$ are evaluated at $t$, rather than at $\tau$.
Then the metric locally (i.e., for $|\vec x| \ll H^{-1}$) becomes
\be \label{localFRW}
ds^2 \simeq - \big[ 1 - \big(\dot H + H^2 \big) |\vec x|^2 \big] dt^2 +
\big[ 1 - \sfrac{1}{2} H^2 |\vec x|^2 \big] d \vec x ^2 \; ,
\ee
which corresponds to Newtonian gauge with potentials
\be
\Phi = -\sfrac{1}{2} \big(\dot H + H^2 \big) |\vec x|^2 \; , \qquad \Psi = \sfrac{1}{4} H^2 |\vec x|^2 \; .
\ee
Corrections to the above metric are of order $H^4 x^4$.
Notice that Newtonian gauge is a complete gauge-fixing only for $\vec k \neq 0$ (see e.g.~ref.~\cite{weinberg3}); on the other hand our potentials are supported at $\vec k =0$ exclusively, $|\vec x|^2 \sim \nabla^2_k \, \delta^{(3)} ( k)$. We will come back to the resulting extra gauge-freedom later.
The above form of the metric, eq.~(\ref{localFRW}) is convenient because of its being locally a small deviation from Minkowski space for all times. That is all along the worldline of the observer sitting at $\vec x=0$, the metric is $\eta_{\mu\nu}$ and the connection vanishes. Therefore our gauge-fixing corresponds to Fermi normal coordinates.

Despite the name, Newtonian gauge does not imply any Newtonian limit, and is still fully relativistic. However, for a matter dominated universe, all the non-trivial cosmological dynamics is captured in our gauge (\ref{localFRW}) by Newtonian local physics. Indeed, the Newtonian potential $\Phi$ above obeys Poisson's equation
\be \label{Poisson}
\nabla^2 \Phi = 4 \pi G \rho  \; ,
\ee
provided that $H^2$ and $\dot H$ satisfy Friedmann equations. 
The local fluid velocity field is given by the Hubble flow,
\be
{\vec v} = H \vec x \; ,
\ee
which solves Newton's law,
\be
\dot {\vec v} = - \vec \nabla \Phi = \frac{\ddot a}{a} \vec x \; .
\ee
The covariant conservation of the stress-energy tensor follows from the Newtonian continuity equation,
\be \label{continuity}
\dot \rho + \vec \nabla \cdot ( \rho \vec v) = \dot \rho + 3 H \rho = 0 \; .
\ee
In fact, although we deduced the local form of a cosmological solution from its exact, global FRW geometry, for a matter-dominated universe one can {\em derive} the correct FRW dynamics just from local Newtonian physics. Indeed, assuming isotropy around the origin, Poisson's equation with constant $\rho$ yields $\Phi = \frac{2\pi G}{3} \rho \, |\vec x| ^2$. The Hubble flow $\vec v \propto \vec x$ is the only velocity field that is homogeneous---it is invariant under the combined action of a translation and a (time-dependent) Galilean boost
\footnote{Notice that,  by performing a time-dependent boost we are in fact going to an accelerated reference frame. Since in Newtonian physics acceleration is physical, strictly speaking the velocity field is not homogenous---the only inertial comoving observer is that sitting at the origin. However, even in Newtonian physics the only observable accelerations are the {\em relative} ones---that is the essence of the (Newtonian) equivalence principle. From this viewpoint all our Newtonian comoving observers are equivalent.
}. 
Newton's law (or Euler's equation) $\dot {\vec v} = - \vec \nabla \Phi$ relates the proportionality factor (i.e., the Hubble parameter $H(t)$) to $\rho$, via a differential equation. The continuity equation (\ref{continuity}) then gives the correct time-dependence of $\rho$. 
The fact that this Newtonian derivation gives the correct GR results is no accident: for non-relativistic sources and at distances much shorter than the curvature radius, the dynamics {\em must} reduce to Newtonian gravitational physics.
On the other hand for a relativistic fluid with sizable pressure, the local interpretation is subtler because of relativistic effects, and the coupling to gravity is not fully captured by the Newtonian potential.

For our purposes it will be more convenient to employ the residual gauge-freedom of Newtonian gauge at $\vec k =0$ and work with slightly different coordinates, where the metric is locally a small {\em conformal} deformation of Minkowski space. The downside is that in such a form the metric is not near-Minkowskian for all times, i.e.~deviations from Minkowski space now will only be small locally in space {\em and} time.  Around $\vec x=0, t=0$ the needed infinitesimal gauge transformation is
\be
\xi_0 = - \sfrac14 \big( 2 \dot H + H^2 \big) \big( |\vec x|^2 t + \sfrac13 t^3 \big)
 \; ,\qquad  
\xi_i =  \sfrac14 \big( 2 \dot H + H^2 \big)  x_i t^2 \; ,
\ee
under which the metric (\ref{localFRW}) transforms to
\be \label{conformal}
ds^2 \simeq \Big[ 1 - \sfrac{1}{2} H^2 |\vec x|^2 + \sfrac12 \big( 2 \dot H + H^2 \big)  t^2 \Big] 
\big( -dt^2 + d\vec x^2  \big ) \; .
\ee
At the order we are working at $H^2$ and $\dot H$ can be evaluated at $t=0$ and treated as constants.
The Newtonian potentials now are
\be \label{phiconformal}
\Phi  = - \sfrac{1}{4} H^2 |\vec x|^2 + \sfrac14 \big( 2 \dot H  +  H^2 \big) t^2 \; , \qquad \Psi = - \Phi \; . 
\ee

\subsection{Mixing with $\pi$}
\label{mixing}

Let us consider a long-distance modification of GR where, at least locally and for weak gravitational fields,  deviations from GR are due to a light scalar degree of freedom $\pi$ kinetically mixed with the metric. 
At quadratic order, this amounts to replacing the Einstein action $\sqrt{-g} \, R$ with $\sqrt{-g} \, (1-2 \pi) R$ plus
whatever dynamics $\pi$ may have on its own. For reasons that will become clear in the following, even for weak gravitational fields and small $\pi$ values we allow for generic non-linearities in the $\pi$ sector, whereas we treat the gravitational field and $\pi$'s contribution to it at linear order (in the e.o.m.).
In this conformal frame we do not couple $\pi$ directly to matter, which we assume to be minimally coupled to the metric $g_{\mu\nu}$. This ensures that introducing $\pi$ does not impair the universality of gravitational interactions.
If we demix $\pi$ and $g_{\mu\nu}$ by performing the Weyl rescaling
\be \label{Weyl1}
h_{\mu\nu} = \hat h_{\mu\nu} +2 \pi \, \eta_{\mu\nu} \; ,
\ee
the action becomes
\be\label{actionpi}
S =  \int \! d^4 x \, \Big[ \sfrac12 \mpl^2 \sqrt{- \hat g} \hat R + \sfrac12  \hat h_{\mu\nu} T^{\mu\nu}  + {\cal L}_\pi + \pi \, T^\mu {}_\mu \Big ] \; ,
\ee
where once again we are working at quadratic order in $\hat h_{\mu\nu}$, and ${\cal L}_\pi$ encodes the so far unspecified dynamics of $\pi$. Now, according to the discussion in the Introduction, if we want the modification of gravity mediated by $\pi$ to be an interesting one, $\pi$'s effect on the geometry perceived by matter should be dominated by the coupling $\pi \, T^\mu {}_\mu$ rather than by $\pi$'s contribution to the stress-energy tensor sourcing $\hat h_{\mu\nu}$. For this reason, in this conformal frame we neglect the gravitational backreaction associated with $\pi$ (we will come back to the validity of this approximation in sect.~\ref{gravity}). In this case, the field equations for $\hat h_{\mu\nu}$ are the same as in pure GR, and so for given sources and boundary conditions, their solution will be the GR one. This means that if we know the `physical' metric $h_{\mu\nu}$ for given $T_{\mu\nu}$ in our modified theory, we can get the $\pi$ configuration associated with it by subtracting the corresponding GR solution $\hat h_{\mu\nu}$ with the same $T_{\mu\nu}$, via eq~(\ref{Weyl1}).
However in comparing the two metrics we must allow for gauge transformations, because the Weyl rescaling  eq~(\ref{Weyl1})  does not preserve a generic gauge-fixing. The first step is to fix gauges for $h_{\mu\nu}$ and $\hat h_{\mu\nu}$ such that their difference is conformally flat (if this is not possible, it means that the modification of gravity under study is {\em not} due to a scalar kinetically mixed with the metric). Then
\be \label{pi_conformal}
\pi \, \eta_{\mu\nu} = \sfrac12 (h_{\mu\nu} - \hat h_{\mu\nu})_{\rm conformal} \; .
\ee
However such a gauge choice is not unique. Namely, we are still free to perform infinitesimal conformal transformations, under which the metric changes by a conformal factor. Infinitesimal dilations shift $\pi$ by a constant, whereas infinitesimal special conformal transformations add a linear piece  to it,
\be
\pi \to \pi + c + b_\mu x^\mu \; .
\label{ambiguity}
\ee
This means that without further knowledge of the $\pi$ dynamics, $\pi$ is unambiguously determined by the geometry only up to a constant and a linear piece. Alternatively, we could try to determine $\pi$ by comparing the curvature tensors, which characterize the geometry in a gauge-invariant fashion. However at linear order we have
\be
R_{\mu\nu} = \hat R_{\mu\nu} - 2 \di_\mu \di_{\nu} \pi - \eta_{\mu\nu }\Box \pi \; ,
\ee
which is only sensitive to the second derivatives of $\pi$. Once again, the constant and linear pieces in $\pi$ are left undetermined.

Consider now a cosmological solution in such a modified-gravity theory. According to our discussion of sect.~\ref{cosmology}, given the solution $a(\tau)$ we can choose coordinates such that locally around $\vec x=0, t=0$ the metric is a small conformal perturbation of Minkowski space, eq.~(\ref{conformal}). If we were to do same in GR, at the same value of $\rho$ and $p$ we would get a different conformal factor, with $H_0$ and $\dot H_0$ in place of $H$ and $\dot H$.
Then, to extract the $\pi$ configuration that is locally responsible for the modified cosmological behavior, we just have to use eqs.~(\ref{conformal}, \ref{pi_conformal}). In terms of the Newtonian potential in conformal gauge
\bea\label{extractpi} 
\pi(\vec x, t) &  = & \Phi - \hat \Phi \\
 & = & - \sfrac{1}{4} (H^2 - H_0^2) |\vec x|^2 + \sfrac14 \big[ 2 (\dot H - \dot H_0) +  (H^2 - H_0^2) \big] t^2 
 +  c + b \, t
 \label{generalpi}\; ,
\eea
where the GR quantities $H_0$ and $\dot H_0$ are to be computed at the same values of $\rho$ and $p$ as $H$ and $\dot H$.
We have included the undetermined constant and linear pieces discussed above. In particular, given isotropy, the linear piece can only be proportional to $t$.

\subsection{An example: DGP cosmology}\label{DGPcosmo}

As an example of our scenario and of the effectiveness of our method, we consider cosmological solutions in the DGP model. We will see that locally the modifications to the cosmological dynamics are {\em fully} captured by a 4D scalar.
The four-dimensional modified Friedmann equation reads \cite{deffayet}
\be \label{FriedmannDGP}
H^2  = \sfrac1{3 \mpl^2} \rho +  \epsilon \,  m H = H_0^2  + \epsilon \,  m H  \; ,
\ee
where $H_0^2 =  \frac{1}{3 \mpl^2} \rho$ is the Hubble rate we would have in GR at the same value of $\rho$ and $p$, $m = 1/L_{\rm DGP}$ is the critical scale, 
and $\epsilon$ is the sign that distinguishes between the two branches of solutions. We can absorb $\epsilon$ into the sign of $m$. Then, the self-accelerated branch corresponds to positive $m$.  Taking the time derivative of the above equation, and using the covariant conservation of energy, $\dot \rho = - 3 H (\rho+p)$, we get the second Friedmann equation,
\be \label{FriedmannDGP2}
\dot H = - \sfrac1{2\mpl^2} (\rho+p) \cdot  \big(1-  m/2H \big)^{-1}= \dot H_0 
+ \sfrac12  m (\dot H/H)  \; ,
\ee
where once again $\dot H_0 = - \frac{1}{2\mpl^2}(\rho+p)$ is the corresponding GR value for the given sources.
Now we want to interpret the deviations from GR as due, at least locally in space and time, to a scalar field $\pi$ kinetically mixed with the metric. 
Then eq.~(\ref{generalpi}) tells us the solution for $\pi$,
\be \label{piDGP}
\pi (\vec x, t)= - \sfrac14 m H \cdot |\vec x|^2 + \sfrac14 m \big(\dot H/ H + H \big) \cdot t^2 + c + b \, t \; . 
\ee
We claim that this is the brane-bending mode identified in ref.~\cite{LPR}. To see this, consider the e.o.m.~that comes from varying the 4D effective action of ref.~\cite{LPR} with respect to $\pi$,
\footnote{
The present $\pi$ has a different normalization from that of refs.~\cite{LPR, NR}. For instance, with respect to ref.~\cite{NR}
\be
\frac {\hat \pi}{M_4} \Big|_{\rm there}=  2  \cdot \pi _{\rm here} \; , \qquad M_4 ^{\rm there}= \sfrac 12 \mpl^{\rm here} \; .
\ee
}
\be
6 \Box \pi - \frac{4}{m^2} (\di_\mu \di_\nu \pi)^2 + \frac{4}{m^2} (\Box \pi)^2 = -T^\mu {}_\mu = \rho - 3p \; .
\ee
If we plug the $\pi$ configuration (\ref{piDGP}) into this e.o.m., we get a linear combination of the two modified Friedmann equations above, namely $12\times(\ref{FriedmannDGP})+6 \times (\ref{FriedmannDGP2})$.
This shows that, despite being intrinsically a five-dimensional model, DGP is well described at short distances by a local 4D theory of a scalar kinetically mixed with the metric \cite{LPR, NR}. This description is locally {\em exact}, with no need of subtle `decoupling' limits---it yields the correct modification to the Friedmann equation as implied by the full 5D theory.


\subsection{The structure of ${\cal L}_\pi$}\label{structure}
We should now characterize the general structure of ${\cal L}_\pi$ that corresponds to a potentially realistic modification of gravity.
On one side we want $\pi$ to give rise to an $O(1)$ modification of the Hubble flow, meaning that $\pi$ exchange has gravitational strength on cosmological scales. On the other we cannot tolerate deviations larger than $O(10^{-3})$ from Einsteinian gravity at solar system distances. It is obviously impossible to satisfy these requests when the $\pi$ dynamics is linear throughout the universe. However if non-linearities are  important, then the extrapolation between different length scales is non-trivial and it would be in principle possible to have a viable theory. The DGP model with non linearities corresponding to the Vainshtein effect represents an interesting example in that sense. So a first necessary request we want to make is that, even in the assumed regime where $\pi$ can be treated as a small perturbation of the geometry, its dynamics is nonetheless non-linear.
 Consider then the general cosmological solution for $\pi$.  At subhorizon distances the leading behaviour is  captured by a quadratic approximation 
\begin{equation}
\pi_c=C+ B_\mu x^\mu+A_{\mu\nu}x^\mu x^\nu+O(x^3H^3)\,.
\label{quadratic}
\end{equation}
We remark that the quadratic approximation to the solution is invariant under the combined action of a spacetime translation and a shift $\pi(x)\to\pi(x)+b_\mu x^\mu+c$. More precisely
\begin{equation}
\pi_c(x)=\pi_c(x+\delta)-2A_{\mu\nu}\delta^\mu x^\nu-B_\mu\delta^\mu\, ,
\label{invariance}
\end{equation}
that is $b_\mu=-2A_{\mu\nu}\delta^\nu$ and $c=-B_\mu\delta^\mu$. Barring accidents, a symmetry in a solution follows from an invariance of the equations of motion and thus constrains the form of the Lagrangian. In the case at hand, taking for granted translational invariance, we would deduce that the lagrangian we want must be invariant under the shift 
\footnote{Strictly speaking, for a given a solution, characterized by $A_{\mu\nu}$ and $B_\mu$, combining the transformation in eq.~(\ref{invariance})  with a translation we obtain a  4-parameter  family of transformations where $b_\mu$ and $c$ are functions of a 4-vector $\delta $. However the commutator of these tranformations with translations is a constant shift $\pi \to \pi +c$, so we recover the full 5 parameter group.}
\be \label{shift}
\pi(x)\to\pi(x)+b_\mu x^\mu+c \; .
\ee
However before taking this deductive step, we have to be more careful, as our  symmetry just trivially follows from our quadratic approximation: the solution to any scalar field theory
can be approximated by its Taylor expansion at second order in the coordinates around one point! In fact, normally, approximating the solution to quadratic order in the coordinates corresponds to linearizing the equations of motion in the field. This is easily seen for an ordinary two derivative lagrangian $-\frac{1}{2}(\partial\pi)^2-V(\pi)$ where the  equation of motion  linearized in $\varphi=\pi(x)-\pi(0)$ is
\begin{equation}
\Box\varphi= V'(\pi(0))={\rm {const}}\,.
\label{linearized}
\end{equation}
Notice indeed that the above equation  admits eq.~(\ref{quadratic}) among its solutions. Moreover  eq.~(\ref{linearized}) is, as it should be expected, invariant under the shift $\varphi(x)\to\varphi(x)+b_\mu x^\mu+c$. 
Therefore in ordinary cases the symmetry $\pi(x)\to\pi(x)+b_\mu x^\mu+c$ of the approximated solution corresponds trivially to an  accidental symmetry of the linearized equations of motion.
However, as explained above, we are here interested in the cases where the $\pi$ dynamics, even at local quadratic order in the coordinates, is described by a non-linear equation of motion. As follows from  our discussion we are thus lead to make a second non trivial assumption, namely that the equations of motion of $\pi$ be invariant under the shift symmetry (\ref{shift}). This is equivalent to demanding that each $\pi$ in the equation of motion be acted upon by at least two derivatives. In fact in order to avoid trouble with ghost degrees of freedom, a  third final request we should make is that the $\pi$ equation of motion be just of second order. Again the need for this request is exemplified  by comparing the DGP model to Fierz-Pauli massive gravity. In the latter, unlike the first, the analogue of the $\pi$ dynamics leads to 4th order equations \cite{Creminelli:2005qk}, with a resulting ghost at the onset of non-linearity. Our requests imply the equation of motion to take the form
 \begin{equation}
 \frac{\delta{\cal L}_\pi}{\delta \pi}=F(\partial_\mu\partial_\nu\pi)
 \label{variation}
 \end{equation}
 where $F$ is a non-linear Lorentz invariant function of  the tensor
$ \partial_\mu\partial_\nu\pi$. In the next paragraphs we shall classify the
Lorentz invariants satisfying eq.~(\ref{variation}). It turns out that there is only a handful of invariants (more precisely 5 of them  in 4D) that do so. Such very constrained  class of Lagrangians represents a generalization of the $\pi$ Lagrangian describing modifications of gravity at sub-Hubble distances in the DGP model \cite{LPR}. This is indeed not surprising, since our three constraints correspond to the three main properties that guarantee viability of the DGP model at solar system  distances. To conclude, the shift corresponding to $c$  makes $\pi$ a Goldstone boson. However the vectorial parameter $b_\mu$ corresponds to constant shift of the gradient $\partial_\mu \pi\to\partial_\mu \pi+ b_\mu$. That 
is the space-time generalization of the Galilean symmetry $\dot x\to \dot x+v$ of non-relativistic mechanics (0+1 field theory). We are thus naturally led to term $\pi\to\pi+bx$ Galilean transformation and to call $\pi$ {\it {galileon}}.

Before proceeding we would like to mention that Galilean invariance of ${\cal L}_\pi$ can be motivated in an alternative  and perhaps more direct way. Indeed as discussed in section \ref{mixing} the modification brought by $\pi$ to the geometry is invariant
under the change in eq.~(\ref{ambiguity}). That is to say that the correction to the Ricci (and Riemann) tensor due to $\pi$,  $ \Delta R_{\mu\nu}=-2\partial_\mu\partial_\nu \pi-\eta_{\mu\nu} \Box \pi$,  is unaffected by eq.~(\ref{ambiguity}).  Now, a potentially realistic modification of gravity can be defined as one that corrects curvature at $O(H^2)$ at cosmological scales but which is suppressed with respect to ordinary GR at shorter scales where the curvature due to local sources is bigger than $H^2$.  As we said, such a behaviour may in principle arise provided the $\pi$ dynamics is non-linear.  Then if we further assume the $\pi$ equations of motion only depend on $\Delta R_{\mu\nu}$ 
\begin{equation}
F(\Delta R_{\mu\nu})=-T^\mu {}_\mu
\end{equation}
we have that corrections to the geometry are directly determined by the distribution of mass in the universe  (and by  boundary conditions). In this situation, we have a better chance to have a robust implementation of the Vainshtein effect where $\Delta R_{\mu\nu}$ is screened sufficiently near localized sources. Thus one viewpoint on  Galilean invariance   is that it  is needed to have a robust and economic implementation of the Vainshtein effect.
We are thus led to classify the Lorentz invariant operators satisfying eq.~(\ref{variation}). These Lagrangian terms must necessarily involve $n$ powers of $\pi$ and $2n-2$ spacetime derivatives. Of course the generic term of this form is not Galilean invariant  and correspondingly its contribution to the e.o.m.~involves also first derivatives of $\pi$. It thus requires some work to find these peculiar invariants at each order in $n$.  The simplest possibility is given by $n=1$, where we have the `tadpole' ${\cal L}_1=\pi$. Notice that the tadpole is invariant under Galilean transformations only up to a total derivative $\delta{\cal L}_1=b_\rho x^\rho+c=\partial_\mu(b_\rho x^\rho x^\mu/5+c x^\mu/4)$. But this ensures anyway invariance of the equations of motion and conservation of a local current. The next to simplest example
 is the standard kinetic energy, 
$(\di \pi)^2$, whose variation under a Galileo boost is
\be 
\delta \, (\di \pi)^2 = -4 c^\mu \, \di_\mu \pi \; ,
\ee
which is a total derivative. A less trivial example is the DGP cubic interaction of ref.~\cite{LPR}, $\Box \pi \, (\di \pi)^2$, whose variation is
\be \label{deltaDGP}
\delta \big( \Box \pi \, (\di \pi)^2 \big)= -4 c^\mu \, \di_\mu \pi \Box \pi =
- 4 c^\mu \, \di^\alpha \big[  \di_\alpha \pi \di_\mu \pi - \sfrac12 \eta_{\alpha\mu} (\di \pi)^2 \big] \; .
\ee
Again, as follows from Galilean invariance, the variation of the above term w.r.t.~$\pi$ only depends on second derivatives of $\pi$. We can go on and study higher order invariants.
In the Appendix we show that formally there exists one and only one such Galilean-invariant at each order in $\pi$. However, since each derivative Lagrangian term we will construct will be associated with one Cayley invariant of the matrix $\di_\mu \di_\nu \pi$, we will have as many non-trivial derivative Galilean invariants as the rank of $\di_\mu \di_\nu \pi$, that is the number of spacetime dimensions. In particular, in 4D we only have five invariants: from the tadpole term, to a quintic derivative interaction
\footnote{In 1D, i.e.~mechanics, we just  have two: the  Galilean kinetic energy $\frac12 m \, \dot x^2$ and the linear potential $x$.}.  

At $n$-th order in $\pi$, the generic structure of a Galileo-invariant Lagrangian term is $(\di^2 \pi)^{n-2} \di \pi \di \pi$. For notational convenience, let's denote by $\Pi$ the matrix of second derivatives of $\pi$, $\Pi^\mu {}_\nu \equiv \di^\mu \di_\nu \pi$. Also the brackets $[ \, \dots ]$ stand for the trace operator, and the `$\cdot$' stands for the standard  Lorentz-invariant contraction of indices. So, for instance
\be
[\Pi] \, \di \pi \cdot \di \pi \equiv \Box \pi \, \di_\mu \pi \di^\mu \pi\; .
\ee
Then, up to fifth order in $\pi$ the Galileo-invariant terms are
\begin{eqnarray} 
{\cal L}_1 & = & \pi \label{lag1}\\
{\cal L}_2 & = & -\sfrac12 \,  \di \pi \cdot \di \pi \\
{\cal L}_3 & = & - \sfrac12 \,  [\Pi] \, \di \pi \cdot \di \pi \\
{\cal L}_4 & = & - \sfrac14 \big( [\Pi]^2 \,  \di \pi \cdot \di \pi - 2 \,  [\Pi] \, \di \pi \cdot \Pi \cdot \di \pi - [\Pi ^2 ] \, \di \pi \cdot \di \pi + 2 \, \di \pi \cdot \Pi ^2 \cdot \di\pi \big) 
\label{Galileo4} \\
{\cal L}_5 & = & -\sfrac15 \big(
[\Pi]^3 \, \di \pi \cdot \di \pi- 3 [\Pi]^2 \,  \di \pi \cdot \Pi \cdot \di \pi
-3 [\Pi] [\Pi^2] \,  \di \pi \cdot \di \pi
+6 [\Pi] \, \di \pi \cdot \Pi^2 \cdot \di\pi
 \nonumber 
 \\
&& 
+2 [\Pi ^3] \, \di \pi \cdot \di \pi
+3 [\Pi ^2] \, \di \pi \cdot \Pi \cdot \di \pi 
- 6 \,  \di \pi \cdot \Pi^3 \cdot \di\pi \big)
\end{eqnarray}
The overall normalizations have been chosen to have simple normalizations in the equations of motion, see below.
Higher order Galileo-invariants are trivial in 4D, being just total derivatives.

For our purposes it is more convenient to work directly at the level of the equations of motion, where there are no integration-by-parts ambiguities, and only second derivatives appear. Defining ${\cal E}_i \equiv \frac{\delta {\cal L}_i}{\delta \pi}$, we get
\begin{eqnarray} 
{\cal E}_1  & = &1 \label{eom1}\\
{\cal E}_2  & = & \Box \pi \label{eom2}\\
{\cal E}_3  & = & (\Box \pi)^2 - (\di_\mu \di_\nu \pi)^2 \\
{\cal E}_4 & = & (\Box \pi)^3 - 3 \,  \Box \pi  (\di_\mu \di_\nu \pi)^2 +2 (\di_\mu \di_\nu \pi)^3  \\
{\cal E}_5 & = & (\Box \pi)^4 -6  (\Box \pi)^2  (\di_\mu \di_\nu \pi)^2 +8 \,  \Box \pi  (\di_\mu \di_\nu \pi)^3 + 3 \big[ (\di_\mu \di_\nu \pi)^2 \big]^2 - 6 (\di_\mu \di_\nu \pi)^4 \label{eom5}
\end{eqnarray}
where by $(\di_\mu \di_\nu \pi)^n$ we denote the cyclic contraction,
$ (\di_\mu \di_\nu \pi)^n \equiv [ \Pi ^n ]$.

The complete Lagrangian for $\pi$ is a linear combination of the above invariants
\be \label{Lsum}
{\cal L}_\pi = \sum_{i=1} ^5 c_i \,  {\cal L}_i \; ,
\ee
where the $c_i$'s are generic coefficients. Likewise, the equation of motion  is
\be  \label{eom}
{\cal E} \equiv \frac{\delta {\cal L}_\pi}{\delta \pi} = \sum_{i=1} ^5 c_i  \,
{\cal E}_i = -T^\mu {}_\mu \; .
\ee


\subsection{Technical analysis}\label{technical}

We now want to investigate the existence of  well-behaved self-accelerating solutions. By `self-accelerating' we mean solutions for $\pi$ that correspond to a deSitter geometry even for vanishing stress-energy tensor. Eq.~(\ref{generalpi}) gives us the relevant $\pi$ configuration, 
\be\label{localdS}
\pi_{\rm dS}(x) = -\sfrac14 H^2 \, x_\mu x^\mu \; , \qquad \di_\mu\di_\nu \pi_{\rm dS} = - \sfrac12 H^2 \eta_{\mu\nu} \; ,
\ee
where we used $H_0 = \dot H_0 = 0$, because the stress-energy tensor vanishes, and $\dot H = 0$, because we are looking for a deSitter solution.
The curvature $H^2$ is determined by the  equation of motion (\ref{eom}), with $T^\mu {}_{\mu} = 0$. Indeed plugging the deSitter ansatz into the e.o.m.~we get an algebraic equation for $H^2$,
\be \label{H2}
 c_1-2 c_2 \, H^2 +3 c_3 \, H^4 -3 c_4 \, H^6 + \sfrac32 c_5 \,  H^8 = 0 \; .
\ee
Then, assuming we have a deSitter solution, we are interested in studying its stability and in general the dynamics of local perturbations that do not change the asymptotics. With an abuse of notation, we will refer to these perturbations by $\pi$. That is, we perform the replacement $\pi \to \pi_{\rm dS} + \pi$ everywhere in the Lagrangian and in the e.o.m. The background deSitter solution (\ref{localdS}) is Lorentz-invariant. Then Lorentz symmetry is unbroken, and it must be an invariance of the action for the perturbations. Furthermore, the background is invariant under the combined action of a spacetime translation and of an internal Galileo-boost,
\be
x^\mu \to x^\mu + a^\mu \; , \qquad \pi_{\rm dS} \to \pi_{\rm dS} + \sfrac12 H^2 a_\mu x^\mu \; ,
\ee
which must also be a symmetry of the perturbations' dynamics.
But in the e.o.m.~only second derivatives appear, and they are automatically invariant under internal Galileo boosts. Also, the background's second derivatives are translationally invariant. This means that the e.o.m.~for the perturbations is separately invariant under spacetime translations and under internal Galileo-boosts, as well as under spacetime Lorentz boosts. That is, apart from boundary terms in the action, the spacetime Poincar\'e symmetry and the internal Galileo boosts are unbroken.
The e.o.m.~for the perturbations then must be  a linear combination of our Galileo-invariants (\ref{eom2}--\ref{eom5}),
\be \label{eomprimed}
\frac{\delta {\cal L}_\pi}{\delta \pi} \bigg|_{\rm dS} = \sum_{i=2} ^5 d_i  \,
{\cal E}_i \; ,
\ee
but now with different parameters $d_i$. The latter are linear combinations of the original parameters $c_i$ with $H^2$-dependent coefficients.  The constant term ${\cal E}_1$ does not appear in (\ref{eomprimed}) because it is precisely the e.o.m. evaluated on the deSitter background (\ref{H2}), which vanishes. Explicit expressions for the mapping $\{c_i \} \leftrightarrow \{d_i \}$ $(i=2,\ldots,5)$ are given in the Appendix. The exact relation will not be important for us. Suffice it to say that it is invertible.

In conclusion, as long as we are interested in solutions with deSitter asymptotics, we can forget about the original Lagrangian coefficients $c_i$, pretend to be in flat space, and parameterize the dynamics of perturbations with arbitrary coefficients $d_i$. The determination of the deSitter Hubble parameter does not impose any constraint on the coefficients $d_i$: for a given set $d_2,\ldots,d_5$ we can compute the corresponding $c_2,\ldots,c_5$ and by choosing appropriately $c_1$ there always exists a solution of (\ref{H2}) with positive $H^2$.


In order for our Galilean model to be a fairly realistic description of our universe, it should pass a number of zero-th
order tests. It should admit a deSitter solution that is stable against small perturbations. Furthermore, about this
deSitter configuration there should exist spherically symmetric solutions that describe the $\pi$ field generated by compact sources. These configurations should also be ghost-free. Finally, we will require that small pertubations about these solutions do not propogate superluminally. As we will see our model fails this last test. We will briefly comment on this in sect.~\ref{discussion}.

For spherically symmetric configurations, $\pi = \pi(r)$, the e.o.m.~drastically simplifies. In particular, like in DGP, it becomes an {\em algebraic} equation for $\pi'(r)$. This is because the $\pi$ e.o.m.~is the divergence of the Noether current associated with the shift symmetry $\pi \to \pi+ c$. For generic $\pi$ configurations the current $J^\mu$ involves first as well as second derivatives of $\pi$. Once we take the divergence, all third derivatives cancel, thus yielding   two-derivative e.o.m. However for a radial configuration $\pi(r)$ the current cannot depend on $\pi''(r)$---otherwise the third derivatives could not cancel in the e.o.m.:
\be
\di_\mu \,  J^\mu(\pi', \pi''; r) = \frac1{r^2} \frac{d}{dr} \big(r^2 J^r \big)
= \frac2{r} J^r + \frac{\di J^r}{\di \pi'} \, \pi''(r) + \frac{\di J^r}{\di \pi''} \, \pi'''(r) + \frac{\di J^r}{\di r}\; .
\ee
Then the current only depends on $\pi'(r)$ (and on $r$), and the e.o.m.~is easily integrated via Gauss's theorem, thus yielding an algebraic equation for $\pi'(r)$. Moreover, since each Galilean term in the e.o.m.~is of the form ${\cal E}_{n+1} \sim (\di \di \pi)^n$, the corresponding piece in the current is schematically $J_{n+1} \sim (\di \di \pi)^{n-1} \di \pi$. For radial configurations $\di \pi \sim \pi'$ and $\di \di \pi \sim \pi'', \frac{\pi'}{r}$, and since the current cannot contain $\pi''$, we have
\be
J_{n+1} \sim  r \cdot \big( \pi'/r \big)^n \; .
\ee
This means that the e.o.m.~reduces in fact to an algebraic equation for $(\pi'/r)$.
Explicitly, we have
\begin{eqnarray}
{\cal E}_2 & \to &  \sfrac{1}{r^2} \sfrac{d}{dr}(r^2 \,\pi') \\
{\cal E}_3 & \to &  \sfrac{2}{r^2} \sfrac{d}{dr}(r \, \pi'^2) \\
{\cal E}_4 & \to &  \sfrac{2}{r^2} \sfrac{d}{dr}(\pi'^3) \\
{\cal E}_5 & \to & 0
\end{eqnarray}
The vanishing of the last contribution is due to the fact that for a time-independent $\pi$ the matrix $\di_\mu \di_\nu \pi$ has rank three, thus the fourth-order invariant identically vanishes. For a source localized at the origin,
$\rho = M \delta^3 (\vec r)$,
\be \label{radialeq}
\frac{\delta {\cal L}_\pi}{\delta \pi} = \sum_i d_i \, {\cal E}_i  =
\frac{1}{r^2}\frac{d}{dr} \, r^3 \left[d_2 \big( \pi'/r \big) + 2 d_3 \big( \pi'/r\big)^2 + 2 d_4\big( \pi'/r\big)^3  \right]
= M \delta^3 (\vec r) \; ,
\ee
we get
\be \label{radial}
d_2 \, \big( \pi'/r \big) + 2 d_3 \,  \big( \pi'/r\big)^2 + 2 d_4 \, \big( \pi'/r\big)^3 = \frac{M}{4\pi r^3} \; .
\ee


\noindent
{\em Existence}

\noindent
We see that the existence of a radial solution for all $r$'s constrains the coefficients $d_i$ to some extent. Indeed varying $r$ from zero to infinity makes the r.h.s.~of eq.~(\ref{radial}) span all positive real numbers. Let's call $P(y)$ the polynomial of $y = \pi'/r$ on the l.h.s.,
\be \label{P}
P(y) \equiv d_2 \, y + 2 d_3 \,  y^2 + 2 d_4 \, y^3  \; .
\ee
Then, for a solution to exist $P(y)$ must be invertible when it is positive. 
We already know that $d_2$ must be strictly positive for the background (deSitter) solution to be stable against small perturbations: $d_2$ is the coefficient of $-\frac12 (\di \pi)^2$ in the perturbations' Lagrangian. Now, far from the source  $y = \pi' / r$ is small, hence the linear term in $P(y)$ dominates. Then, at small $y$'s, positive $P$ implies positive $y$. $P$ is invertible if $P'(y)$ does not cross zero. Thus, in moving to larger $y$'s (smaller $r$'s) $P$ has to increase. As a consequence, $P$ is positive when $y$ is.
This means that the requirement that $P(y)$ be invertible whenever it is positive is equivalent to ask that
\be \label{Pprime}
P'(y) = d_2 + 4 d_3 \, y + 6 d_4 \, y^2> 0 \qquad \forall \,  y>0 \; .
\ee 
A radial solution exists iff this condition is obeyed.
At large $y$ (small $r$), the quadratic term in $P'(y)$ dominates, thus $d_4$ has to be non-negative. Then eq.~(\ref{Pprime}) is satisfied for every $y>0$ if we also have $d_3> - \sqrt{ \frac32 \, d_2 \, d_4}$. 
%
%
In conclusion, there exist spherically simmetric solutions with deSitter asymptotics iff
\be \label{existence}
\left\{ \begin{array}{l}
d_2 > 0 \\
d_4 \ge 0 \\
d_3 > - \sqrt{ \frac32 \, d_2 \, d_4}. 
\end{array}\right.
\ee 

\vspace{0.5cm}

\noindent
{\em Stability}

\noindent
We now move to study the stability of the radial solution $\pi_0(r)$ against small perturbations, $\pi = \pi_0(r) + \varphi$.
Given the symmetries of the solution, the quadratic Lagrangian for small perturbations will have the form
\be
{\cal S}_\varphi = \sfrac12 \int \! d^4x \left [K_{t} (r) (\partial_t \varphi)^2 - K_{r} (r) (\partial_r \varphi)^2  - K_{\Omega} (r) (\partial_\Omega \varphi)^2 \right] \; ,
\label{quadraticradial}
\ee
where $(\partial_\Omega \varphi)^2$ is the angular part of $(\vec \nabla \varphi)^2$, and the kinetic coefficients $K_i$ depend on $r$ through the background radial solution $\pi_0(r)$ and its derivatives.
That fact that the full e.o.m., eq.~(\ref{eomprimed}), only contains second derivatives  guarantees that small perturbations about any solution will have a two-derivative quadratic Lagrangian, despite the formal appearance of many high-derivative terms in the original Lagrangian for $\pi$.
For the solution to be stable against short wavelength perturbations, all the $K_i$'s must be positive (see a related discussion in ref.~\cite{NR}).

Rather than obtaining the $K_i$'s by expanding the full $\pi$-Lagrangian at quadratic order in $\varphi$, it is faster to compute them by expanding the e.o.m.~at linear order,
\be \label{flucteq}
{\cal E}[\pi_0 + \varphi] \to \frac{\delta {\cal S}_\varphi}{\delta \varphi} = -K_t (r)\, \ddot \varphi +
\frac{1}{r^2} \partial_r \big(r^2 K_r (r)\, \partial_r \varphi \big) + K_\Omega (r) \,  \partial^2_\Omega \varphi \; ,
\ee
where $\partial^2_\Omega$ is the angular part of the Laplacian.
In particular, we can get the radial kinetic coefficient from the radial e.o.m. Evidently, comparing eq.~(\ref{radialeq}) with the radial part of eq.~(\ref{flucteq}),
\be
K_r (r)  =  P'(y) \big|_{y=\pi_0'/r} \; ,
\ee
where $P(y)$ is the polynomial defined in eq.~(\ref{P}). The positivity of $K_r$ is thus guaranteed by the existence of the radial solution itself, eq.~(\ref{Pprime}).

For the other $K_i$'s we get 
\begin{eqnarray}
K_t (r) & = & \frac{1}{3 r^2} \frac{d}{dr} \left[ r^3 \big(d_2 + 6 d_3 \, y
+18 d_4  \, y ^2 + 24 d_5 \, y^3 \big)_{y=\pi_0'/r} \right] \label{Kt}
\\
K_\Omega (r) & = & \frac{1}{2 r} \frac{d}{dr} \left[ r^2 \big(d_2 + 4 d_3 \, y
+ 6 d_4  \, y ^2 \big)_{y=\pi_0'/r} \right]   \label{KO}
\end{eqnarray}
These depend on $y$ as well as on its derivative with respect to $r$. Also they explicitly involve $r$. Thus it appears that the question of their positivity is now a differential problem rather than an algebraic one. However we can eliminate $y'(r)$ in favor of $y$ and $r$ through the implicit function theorem: we know that the background solution satifies
\be
F(r, y) \equiv  P(y) - \frac{M}{4 \pi r^3} = 0 \; .
\ee
Then, on the solution
\be
y'(r) = - \frac{\partial_r F}{ \partial_y F} \; .
\ee
This yields
\begin{eqnarray}
K_t & = & \frac{d_2^2 + 4 d_2 d_3 \, y +12 (d_3^2 - d_2 d_4) \, y^2
+24 (d_3 d_4 - 2d_2 d_5 ) \, y^3 + 12 (3 d_4^2 - 4 d_3 d_5) \, y^4}
{d_2 + 4 d_3 \, y + 6 d_4 \, y^2}
\\
K_\Omega & = & \frac{d_2^2 +2 d_2 d_3 \, y + (4 d_3^2 - 6 d_2 d_4) \, y^2}
{d_2 + 4 d_3 \, y + 6 d_4 \, y^2}
\label{omega}
\end{eqnarray}
Notice that the explicit $r$-dependence canceled out. This is because both the background e.o.m. and the kinetic coefficients $K_i$ for the fluctuation only involve the matrix of second derivatives of $\pi$. For radial configurations we have $\di \di \pi \sim \pi'', \frac{\pi'}{r}$. Moreover the background e.o.m.~are {\em linear} in $\pi''$, being the divergence of a current only involving $\pi'$. As a consequence on the solution $\pi''$ is a {\em rational} function of $\pi'/r$, which makes the kinetic coefficients $K_i$ rational functions of $\pi'/r \equiv y$.

Recall that the solution spans all positive values of $y$. Thus, we have to require that $K_t$ and $K_\Omega$ are positive for all positive $y$'s.
Given the constraints (\ref{existence}), the denominators are automatically positive. The numerator in $K_{\Omega}$ then is positive for 
$d_3 \ge \sqrt{\frac{3}{2} \, d_2 \, d_4}$  which makes the numerator in $K_t$ also positive provided that $d_5\le \frac34 \frac{d_4^2}{d_3}$.
In summary, the radial solution with deSitter asymptotics exists and is stable against small perturbations iff
\be \label{stability}
\left\{ \begin{array}{l}
d_2 > 0 \\
d_4 \ge 0 \\
d_3 \ge  \sqrt{ \frac{3}{2} \, d_2 \, d_4} \\
d_5 \le \frac34 \frac{d_4^2}{d_3}
\end{array}\right.
\ee

\noindent
{\em Subluminality}

\noindent
Increasing further our level of pickiness, we can ask whether small fluctuations of our solution propagate at subluminal speeds.
The speed of fluctuations propagating radially is simply
\be
c_r^2 = \frac{K_r}{K_t} \; .
\ee
Given the constraints imposed by existence and stability, eq.~(\ref{stability}), this is forced to be {\em superluminal} at large distances from the source  (small $y$). Indeed
\be
c_r^2 = 1 + 4 \frac{d_3}{d_2} y + {\cal O}(y^2) > 1 \; .
\ee
Notice that the sign of the deviation from the speed of light is positive because of the requirement of existence and stability of spherical solutions, independently from the presence or not of the self-accelerating background.   
In moving to smaller distances (larger $y$) the propagation speed can be made subluminal. This can be easily seen from the difference
\be
K_r - K_t = 4 d_2 d_3 \, y + 4 (d_3^2 + 6 d_2 d_4) \, y^2 + 24 (d_3 d_4 + 2 d_2 d_5) \, y^3 + 48 d_3 d_5 \, y^4 \; , 
\ee
which becomes negative at large $y$ if $d_5$ is negative.

On the other hand, angular excitations are always subluminal: their propagation speed is
\be
c_\Omega^2 = \frac{K_\Omega}{K_t} \; .
\ee
The difference between the numerator and the denominator is, apart from an overall positive factor
\be
K_\Omega - K_t \propto - 2 d_2 d_3 \, y - (8 d_3^2 - 6 d_2 d_4) \, y^2 - 24 (d_3 d_4 - 2 d_2 d_5) \, y^3 -12 (3 d_4^2 -4 d_3 d_5) \, y^4 \; , 
\ee
which is negative, as each power of $y$ has  a negative coefficient.
We end up with the following constraints:
\be \label{finalconst}
\left\{ \begin{array}{l}
d_2 > 0 \\
d_4 \ge 0 \\
d_3 \ge  \sqrt{ \frac{3}{2} \, d_2 \, d_4} \\
d_5 < 0
\end{array}\right.
\ee


\section{Some phenomenology}

In this section we will broadly discuss the scales associated with classical non-linearities and with quantum effects.
The analysis can be straighforwardly applied also to the fully conformally invariant case we discuss in sect.~\ref{IR}.

\subsection{The coupling to matter}\label{matter}

The galileon Lagrangian has the general form
\begin{equation}
{\cal L}= f^2\partial \pi \partial \pi F(\partial\partial \pi/H_0^2)
\label{pisquiggly}
\end{equation}
We have directly chosen the scale controlling  higher orders in $\partial \partial \pi$ to be the present Hubble parameter $H_0$,
so that the $\pi$ contribution affects at $O(1)$ the present visible expansion.
$f^2$ is an overall coefficient that gives the $\pi$ Lagrangian the correct dimensions. We can think of $f$ as measuring the coupling between $\pi$ quanta and matter. Indeed in terms of the canonically normalized $\pi_c$, upon demixing at linear order, the coupling to matter is  $\sim \frac{1}{f} \, \pi_c \, T^\mu {}_\mu$. The scale $f$ is bounded by considering the effects, direct and indirect, of $\pi$ on the geometry at scales shorter than or comparable  to $1/H_0$.  
An upper bound is determined by considering the gravitational backreaction to the $\pi$ energy momentum tensor. At subhorizon scales we have $\pi \sim H_0^2 x^2$, so that eq.~(\ref{pisquiggly}) gives $\Delta T_{\mu\nu}=O(f^2H_0^4 x^2)$.
Requesting that $\Delta T_{\mu\nu}$ does not lead to a curvature $\gg H_0^2$ at scales just below $H_0^{-1}$ implies $f\lsim \mpl$.
A comparable lower bound is instead determined by considering the direct contribution of $\pi$ to the geometry perceived by matter. A bound is thus provided by those gravitationally bound systems where the adequacy of GR has been directly tested 
to work at least at $O(1)$. To derive a constraint a qualitative estimate of the effect of $\pi$ is sufficient. To do so we must
figure out which regime, linear or non-linear, characterizes the  dynamics of the perturbation $\pi$ over the cosmological background $\pi_{\rm{ds}}$.
 Very much like in the DGP model, and as discussed in section 1.5, $\pi$ can be treated  as linear whenever  $\partial\partial \pi \lsim O(H_0^2)$, which is normally far enough from a localized gravitational source; while   $\partial\partial \pi \gsim O(H_0^2)$ corresponds to the non-linear Vainshtein regime, that normally sets  in near enough to sources. To get a qualitative idea
 consider a source of total mass $M$ corresponding to a  Newtonian potential $h_N\sim\frac{1}{M_P^2}M/r=R_S/r$. At large enough distance where the linear regime applies the
galileon contribution to the metric goes like $\pi \sim \frac{1}{f^2} M/r=(M_P/f)^2 R_S/r$ so we have
\be
\frac{\partial \partial \pi}{H_0^2}\sim \frac{M_P^2}{f^2} \frac{R_S}{r^3}\frac{1}{H_0^2}= \frac{M_P^2}{f^2} \frac{\cal R}{H_0^2}
\ee
where by ${\cal R}$ we indicate the Riemann curvature induced by the source. Now, a system behaves as gravitationally bound when the local gravitational attraction wins over the tidal force associated with Hubble expansion, that is when ${\cal R} \gtrsim H_0^2$. 
This equation shows that when $f \lesssim M_P$, for any gravitationally bound system $\pi$ is in the Vainshtein regime. To obtain a bound on the case $f \lesssim M_P$ we thus have to estimate the effects of $\pi$ by using the behaviour in the Vainshtein region. From equation (\ref{radial}) we see that at short distances the quartic coupling dominates the equation of motion. The coefficients $d_n$ according to equation (77) scale like 
\begin{equation}
d_n=f^2 \frac{a_n}{H_0^{2n-4}}
\end{equation}
with $a_n \sim {\cal O}(1)$, then the asymptotic behavior in the Vainshtein region is $\pi \sim (M H_0^4/f^2)^{1/3} \, r$.
 The relative size of the $\pi$-potential and the Newtonian one is
\be
\frac{\pi}{h_N}\sim \left (\frac{M_P}{f}\right )^{2/3}\frac {H_0^{4/3} r^2}{R_S^{2/3}}=\left (\frac{M_P}{f}\frac{H_0^2}{\cal R}\, \right )^{2/3} .
\ee
Although the strongest bound on $\pi /h_N$ comes from tests of gravity within the solar system, that bound does not translate into the strongest bound on $f$. This is because ${\cal R}$ within the solar system is much bigger than $H_0^2$.
The strongest bound on $f$ comes instead from more loosely bound systems like galaxy superclusters, at whose scales  ${\cal R}$ is just a bit bigger than $H_0^2$, and for which  general relativity seems valid at $O(1)$ or better. This translates into a lower bound roughly
\be
f> \frac{M_P}{\rm a \,few}
\ee
In the end we basically have no choice other than $f\sim M_P$. 
The fact that both the upper and lower bound on $f$ almost coincide is yet another illustration of how difficult is to come up with consistent modifications of gravity. Notice on the other hand that in the DGP model, where $\pi$ originates from 
the 5D geometry, one automatically obtains $f=M_P$.

\subsection{Strong interaction scales and retardation effects}\label{quantum}

With the scale $f$ basically constrained to be $M_P$ we can now discuss the issue of quantum corrections to the classical picture we developed so far. Our discussion  essentially follows section 7 of ref. \cite{NR}, where the same issue was studied in the case of the DGP model. However, as we shall see, some important differences arise between the two cases.
Using $f=M_P$ in eq.~(\ref{pisquiggly}) and going to the canonical field $\hat \pi\equiv \pi/M_P$,
we have  
\begin{equation}
{\cal L}= \partial \hat\pi \partial \hat \pi F(\partial\partial \hat \pi/\Lambda^3)\qquad\qquad \Lambda^3= \mpl H_0^2 \; ,
\label{picanonical}
\end{equation}
corresponding to a `naive' strong interaction scale $\Lambda\sim (1000 \: {\rm km})^{-1}$ precisely like in DGP.  Here we say `naive' because, while  this result applies around flat space, we expect modifications around non trivial solutions. This is because 
in the Vainshtein region the propagation of the excitations $\varphi=\pi-\pi_{0}$ is dominated by the higher order terms in the lagrangian, thus invalidating the above estimate. In order to understand what happens we shall focus on the spherical solution that was discussed in section 1.5.
 In the case of DGP it was shown in ref. \cite{NR} that, for a spherical solution, inside the Vainstein region   $\varphi=\pi-\pi_{0}$ 
 acquires a large kinetic term. This damps $\varphi$'s quantum fluctuations and raises the effective quantum cut-off  to
 $\tilde \Lambda =\Lambda (R_V/r)^{3/4}$, where the Vainshtein radius of a spherical source is $R_V^3=R_S/H_0^2$. For instance, on the surface of earth one has $\tilde \Lambda^{-1}\sim 1\,{\rm cm}$. 
 From the discussion in section \ref{matter} we have 
 \be
 \pi_0\sim \frac{R_S}{R_V^2} r\qquad\longrightarrow \qquad \tilde y\equiv \frac{1}{H_0^2}\frac{\pi_0'}{r}\sim \frac{R_V}{r}\gg 1 \; ,
 \ee
 $\tilde y$ is the parameter controlling the magnitude of non-linear effects inside the Vainshtein region.
 For instance, from the discussion after eq.~(\ref{quadraticradial}), we have that the kinetic coefficients in the radial and temporal
 directions $K_r$ and $K_t$ generically grow like $\tilde y^2$. On the other hand $K_\Omega$ generically goes like a constant, indicating that angular fluctuations are not generically damped when approaching the source.
Indeed from eq.~(\ref{omega}) we see that this constant behaviour arises when $d_4\not=0$, while in DGP $d_4=d_5=0$ and $K_r,  \, K_t, \, K_\Omega \, \sim \tilde y \gg 1$.
As we shall see in a moment the lack of damping for angular fluctuations is an unpleasant novelty compared to DGP. In order to study how  the fluctuation $\varphi$ self-interacts,  it is useful to write the effective action  by distinguishing $x_\perp$, the spacial coordinates tranverse to $\vec \nabla \pi_0$ (the angular coordinates  for our solution)   and  $x_\parallel=(t,r)$. 
Since strong coupling in our derivatively coupled theory is a short distance phenomenon, in order to
 assess what the effective strong scale is at a given point, it is enough to zoom in on that point and `linearize' our coordinates and treat them as cartesian.
Moreover, precisely as we did with the quadratic lagrangian for $\varphi$, we can work out the interaction terms
working directly on the equations of motion. Qualitatively (that is without paying attention to the different  contractions of indices)
the structure of the effective action is
\begin{eqnarray}
S_{\rm eff} & = & \int d^2 x_\parallel d^2 x_\perp \Bigl[\tilde y^2(\partial_\parallel \varphi)^2+ (\partial_\perp \varphi)^2+ a_3 \frac{1}{\Lambda^3}
(\partial_\perp \varphi)^2\partial_\perp^2\varphi \nonumber \\
&& \qquad + \, a_4 \frac{\tilde y}{\Lambda^3} (\partial_\perp \varphi)^2\partial_\parallel^2\varphi + a_5 \frac{\tilde y^2}{\Lambda^3} (\partial_\parallel \varphi)^2\partial_\parallel^2\varphi + \ldots\Bigr ] \; ,
\end{eqnarray}
where the dots indicate less important interactions of cubic, and higher, order. Notice that the interactions involving only
transverse derivatives are not enhanced by $\tilde y$ factors. This arises  because for a linear solution $\pi_0\propto r$ one has  $\partial_i\partial_j \pi_0\propto\delta_{ij}^\perp $ with $\delta^\perp_{ij}$ representing the metric in the transverse directions. Then if we focus on terms involving only transverse derivatives we have that both the background $\partial_i \partial_j \pi_0$ and the fluctuation are rank two matrices living in the transverse space. But such rank two object vanish beyond cubic order because of the rank condition on our Galilean invariants. To estimate the strong scale 
at a given point, we can treat $\tilde y$ as constant. The difficulty with the above lagrangian is the lack of isotropy. One way to proceed is to change variables and make the kinetic term canonical. This can be achieved by working with rescaled coordinates along the transverse direction
\be
x_\perp =\tilde x_\perp/\tilde y\qquad\longrightarrow \qquad\partial_\perp=\tilde y \,  \tilde\partial_\perp\,, \qquad d^2x_\perp =\frac{d^2\tilde x_\perp}{\tilde y^2} \; ,
\ee
so that the action is rewritten as
\begin{eqnarray}
S_{\rm eff} & = & \int d^2 x_\parallel d^2 \tilde x_\perp \Bigl[(\partial_\parallel \varphi)^2+ (\tilde \partial_\perp \varphi)^2+ a_3 \frac{\tilde y^2}{\Lambda^3}
(\tilde \partial_\perp \varphi)^2\tilde \partial_\perp^2\varphi \nonumber \\
&&\qquad + \, a_4 \frac{\tilde y}{\Lambda^3} (\tilde \partial_\perp \varphi)^2\partial_\parallel^2\varphi + a_5 \frac{1}{\Lambda^3} (\partial_\parallel \varphi)^2\partial_\parallel^2\varphi+\dots\Bigr ]
\; .
\end{eqnarray}
In this form we can easily estimate the scattering amplitudes $2\to 2$ that arise working at second order in the cubic
interaction terms. The most dangerous terms are the first two. The third term gives the same cut off as around flat space,
that is $\sim \Lambda$. The first cubic interaction, the one involving only transverse derivatives, gives an amplitude that roughly scales like
\be
{\cal A}_{2\to 2}\sim  \frac{\tilde y^4}{\Lambda^6}\frac{\tilde p_\perp^8}{{\rm max}(\tilde p_\perp^2,p_\parallel^2)}=
 \frac{1}{\tilde y^4\Lambda^6}\frac{p_\perp^8}{{\rm max}(p_\perp^2/\tilde y^2,p_\parallel^2)} \; ,
\ee
where the $1/p^2$ factor is an estimate of the propagator and where we have rewritten the amplitude also in terms of the unrescaled momentum along the transverse directions.
Notice that we have the dispersion relation $\omega \equiv p_\parallel^0=\sqrt { (p_\parallel^1)^2+p_\perp^2/\tilde y^2} \gsim p_\perp/\tilde y$. The above amplitude is clearly maximized by scattering waves moving along the transverse directions $p_\parallel \sim p_\perp/\tilde y$.  In that case  we deduce an effective momentum cut-off 
\be
\Lambda_{p}\sim \tilde y^{1/3}\Lambda=\left(\frac{R_V}{r}\right)^{1/3}\Lambda\,.
\ee
However because of the anisotropic dispersion relation the corresponding energy cut-off is much lower
\be
\Lambda_{E}\sim \tilde y^{-2/3}\Lambda=\left(\frac{R_V}{r}\right)^{-2/3}\Lambda\,.
\ee
One similarly estimates the amplitude determined by just the second cubic interation term, the one involving both transverse and longitudinal derivatives
\be
{\cal A}_{2\to 2}\sim  \frac{\tilde y^2}{\Lambda^6}\frac{\tilde p_\perp^4\tilde p_\parallel^4}{{\rm max}(\tilde p_\perp^2,p_\parallel^2)}=
 \frac{1}{\tilde y^2\Lambda^6}\frac{p_\perp^4 p_\parallel^4}{{\rm max}(p_\perp^2/\tilde y^2,p_\parallel^2)}
\; .
\ee
With the above amplitude it is convenient to consider two regions: i) $p_\parallel \sim p_\perp/\tilde y$ and ii) $p_\parallel > p_\perp/\tilde y$. In the first region we find the same $\Lambda_p\sim \tilde y^{2/3} \Lambda$ and $\Lambda_E\sim \tilde y^{-1/3}\Lambda$, while in the second region we find that the cut-off is $\sim \tilde y^{1/3}\Lambda$ for both momentum and energy. Finally the third cubic  interaction term, the one involving only parallel derivatives, implies a cut-off $\sim \Lambda$ in both momentum and energy. The combined result of this discussion is therefore that the space cut-off remains the same $\sim \Lambda$ but the energy cut off is significantly lowered
due to the scattering of the slow moving modes along the transverse directions. To get a rough idea we can for instance apply the above results to the earth-moon system, neglecting all other sources of $\pi$ other than the earth.
For the earth we have $R_V\sim 10^{19}$ cm and the radius of the moon orbit around the earth is $R_M\sim 4\times 10^{10}$ cm and thus on length scales of the order of $R_M$ the energy cut-off in time units is
\be
T_E =\Lambda_E^{-1}=\Lambda^{-1} \left (\frac{R_V}{R_M}\right )^{2/3}\sim 10^3\, {\rm s}\,.
\ee
The domain of applicability of our theory is therefore rather limited. Anyway, if the one we just illustrated were the only limitation
we would still, at least in principle, be able to defend the estimate of the effects of $\pi$ on the moon orbit, which was one of the interesting features in DGP. After all in that case the time scale, a month, is much bigger than $T_E$. However there are worse limitations (or better complications) in computing the effects of $\pi$ within the solar system already at the classical level. This is because of the big retardation effects associated with the slow motion of the $\pi$ excitations. In order to compute the $\pi$ field generated by the earth we should first find the $\pi$ solution due to the sun, and then
solve the non linear problem around the earth. Now, given the Vainshtein radius $ R_{V\odot}$ associated with the Sun mass and the earth orbit radius $R_E$  we have that angular fluctuations propagate with a speed
\be
v\sim \frac{R_E}{ R_{V\odot}} \times c\sim 10^{-8} c \; ,
\ee
which is orders of magnitude smaller that the velocity of the earth on its orbit. The computation of the $\pi$ field around the earth could then not be performed in the static limit. Correspondingly Cerenkov radiation in the transverse direction would instead be emitted due to the motion of the earth. The basic conclusion is thus that the lack of enhancement of transverse kinetic terms around a spherical source causes trouble. On one side we have a technical trouble at the classical level,  as huge retardation effects prevent a simple static computation of the background. On the other side the have a conceptual trouble at the quantum level, due to a rather low cut off, worse in energy than in momentum, again because of  retardation. Maybe our conclusion for the solar system are too drastic in that we have not considered other effects, like that due to the galaxy, which will enhance the transverse kinetic terms above their $O(1)$ value we used for our study.
In principle one could try and estimate these effects. For instance by studying the $\pi$ background around two separated spherical sources. The lack of spherical symmetry makes this a much tougher problem that the spherical one we studied, and is beyond the scope of the present paper. 

\subsection{Special cases}

So far we have discussed the most general lagrangian $\mathcal{L}_\pi$ in eq.~(\ref{Lsum}), with all the coefficients $c_i$ different from zero. In the following we examine if restricting to a subclass of these theories can alleviate some of the difficulties found in the previous sections.  

From the qualitative analysis of sec.~\ref{quantum} we can see that the presence of the higher order galilean terms $\mathcal{L}_{4,5}$ is responsible for the slow motion of the $\pi$ excitations that in turn restricts the domain of applicability of the theory as one goes to shorter distances. It's interesting then to consider what happens if we choose $d_4=d_5=0$, or equivalent $c_4=c_5=0$. In this case the only interaction left is the cubic one of the DGP model and the kinetic terms $K_r$, $K_t$ and $K_\Omega$ in the region where classical non-linearities are important have an isotropic enhancement $\propto {\tilde y}$. This large kinetic energy suppresses quantum fluctuations $\varphi$ with the result that the position dependent cutoff increases, for example on the surface of the earth we have $\Lambda_\oplus^{-1} \sim 1$ cm \cite{NR}. The difference however is that ---contrary to DGP--- there can still be a self accelerating solution without ghosts because of the presence in the theory of the tadpole $c_1 \pi$. Moreover the conditions (\ref{stability}) can be satisfied when $d_4=d_5=0$.
\footnote{ Notice in passing that the possibility to choose also $d_3=0$ seems to be compatible with the constraints (\ref{stability}). What goes wrong in this case is that without any non-linear effects in the dynamics of $\pi$ we can have deSitter asymptotics but there are $O(1)$ deviation from GR already at solar system scales. As we stressed throughout the paper, the implementation of the Vainshtein effect is a necessary condition to reconcile theories with a lagrangian of the form (\ref{actionpi}) with present astronomical observations.}

The case $c_1=0$ could be regarded as theoretically preferable, as long as both the Minkowsky and deSitter solutions are stable. For instance in  applications to early cosmology this would potentially allow to describe inflation and Big Bang cosmology within the same effective theory. Moreover for $c_1=0$ the deSitter solution
would arise in the absence of a potential, which is one possible definition of self-acceleration. However one should keep in mind that also the case $c_1\not = 0$ gives a quite novel way of generating a deSitter solution: there is, sure, a potential for the scalar $\pi$ but, unlike the case of a cosmological constant it does not have any, even approximately, stationary points!

Anyway with the above caveat in mind it makes sense to better analize the case $c_1=0$ to find out if our zeroth order requirements can be satisfied.
 Can this case be made compatible also with the existence of a spherically symmetric solution?    
Contrary to the analysis carried out in sect.~\ref{technical}, when $c_1 = 0$ equation (\ref{H2}) gives a constraint on the coefficients $d_i$: we can rewrite it factoring out an overall $H^2$ and defining $Q(\sfrac12 H^2)$ the resulting polynomial:
\be
Q(y) \equiv 4 (d_2 + 3 d_3 \, y + 6 d_4 \, y^2 + 6 d_5 \,  y^3) \; .
\ee
Then, for the background deSitter solution to exist at all, $Q(y)$ must have a zero at positive $y$,
\be \label{Q}
\exists \, y_0 >0 \quad | \quad Q(y_0) = 0 \; . 
\ee
If $d_4=d_5=0$ we can see immediately that (\ref{Q}) cannot be satisfied: it's clearly incompatible with the simultaneous requirement of stable fluctuations around deSitter ($d_2>0$) and existence of a radial solution for all $r$ ($d_3 \ge 0$ (\ref{existence})). This is the well known lesson of DGP. 

However even when $d_4,$ $d_5 \neq 0$ there is some tension between the two algebraic eqs.~(\ref{Pprime}, \ref{Q}). In spite of that, it's easy too see that they can be both satisfied if we add to the conditions (\ref{existence}) the requirement $d_5<0$.  Therefore when $c_1=0$ the presence of the quintic term with a negative coefficient in the Lagrangian becomes crucial to make the deSitter background and spherically symmetric solutions coexist. Notice that $d_5<0$ is also compatible with the requirements of stability and subluminality of the fluctuations (\ref{finalconst}), at least close to the source.

We can still take $d_4=0$ trying to avoid the problem related to the speed of the angular fluctuations.  
In this situation we have $K_r$, $K_\Omega \sim \tilde{y}$ while $K_t \sim \tilde{y}^3$. Even before paying attention to the UV cut-off, this case looks right away uninteresting since the retardation effects are even worse than the general case discussed in the previous section. Indeed all modes have a speed $v\sim c/\tilde y$. Now, from eq.~(\ref{radial}) at $d_4=0$, at a distance $r$ from a spherical source with Vainshtein radius $R_V$ we have $\tilde y\sim (R_V/r)^{3/2}$. So in the case of earth orbiting the sun $v\sim 10^{-12} c$.

\section{The IR completion}\label{IR}

\subsection{Self-acceleration from symmetry}\label{symmetry}

We now consider the issue of whether there exists a ``global completion'' of the local $\pi$ dynamics we have been discussing so far. It may be that at large distances the description in terms of $\pi$  breaks down, like in DGP, where at horizon scales the four-dimensional description is no longer applicable. Here we want to entertain the possibility that instead, the physics at large scales is well described by a four-dimensional Lagrangian for $\pi$. 
To this purpose, it is instructive to characterize the acceleration of the universe in terms of symmetries. If the universe asymptotes to a deSitter phase at late times, it is in fact appoaching an enhanced symmetry configuration. Indeed a generic FRW solution is invariant under spatial translations and rotations, whereas deSitter space is maximally symmetric, with as many isometries as Minkowski space. The cosmological constant is perhaps the most elegant way of realizing such symmetries: having a stress-energy tensor proportional to the metric, it does not introduce any new tensor thus being consistent with the highest possible number of symmetries a metric may have. It is certainly the most economic way, in that it does not introduce any new degree of freedom.
A priori, a possible alternative is the usual way in which symmetries are realized in field theory: at the level of the Lagrangian, {\em off-shell}. That is,
if a solution to the field equations has some symmetries, barring accidents these must be shared by the Lagrangian that gives rise to such a solution. The most economic possibility in this class is to realize the desired symmetries introducing the least possible number of new degrees of freedom. As we will see shortly, for the deSitter symmetry group one scalar suffices. 

Let's start by considering the limit in which gravity is decoupled $M_{\rm Pl} \to \infty$.
Notice that the full Poincar\'e-like symmetry group of four-dimensional deSitter space---the generalization of translations, rotations, and boosts---is $SO(4,1)$, that is, the group of five-dimensional Lorentz transformations. Indeed $D$-dimensional deSitter space can be embedded into $(D+1)$-dimensional Minkowski space as an hyperboloid, which is a Lorentz-invariant submanifold.
This suggests a quick way of constructing 4D theories that realize the deSitter symmetry group as an internal symmetry: write down the most generic effective Lagrangian for a 4D brane living in 5D Minkowksi space. The presence of the brane itself will necessarily break spontaneously some combinations of 5D translations, rotations, and boosts. The two largest possible residual groups are the four-dimensional deSitter group $SO(4,1)$---when the brane is an hyperboloid---and its flat limit, the four-dimensional Poincar\'e group. The existence of deSitter solutions will ultimately depend on the Lagrangian, but may be generically expected given the symmetry arguments above. From this geometrical perspective, it is clear that we are introducing just one new 4D scalar---the position of the brane. Notice that in this approach the fifth ``extra-dimension'' is fictitious---no field propagates in it. It is simply a useful geometric picture of the new degree of freedom. However, if we take it more seriously, we introduce gravity in the 5D ambient space, and we also gauge gravity on the brane, what we get is nothing but the DGP model
\footnote{For DGP an orbifold projection is also necessary.}.
From this viewpoint it is not surprising that DGP admits ``self-accelerating'' deSitter solutions.

Another possibility, which is the one we will analyze in the following, is to make the 4D Lagrangian conformal invariant. The 4D conformal group is $SO(4,2)$, whose maximal subgroups are $SO (3,2)$, four-dimensional Poincar\'e, and $SO(4,1)$---the isometry groups of four-dimensional Anti-deSitter, Minkowski, and deSitter spaces, respectively. It is well known that the spontaneous breakdown of the conformal group to any of these subgroups can be achieved by a single scalar, the dilaton \cite{fubini}. Indeed the 4D conformal group is the isometry group of 5D AdS space, and the dilaton can be thought of as the position of a 4D brane in 5D AdS. Then each of the above breaking patterns corresponds to a maximally symmetric brane geometry. 

Let us stress again that for the moment we are decoupling gravity. We will see in sect.~\ref{gravity} that reintroducing dynamical gravity drastically changes the picture at large distances. However for the local analysis we are carrying out, at distances much smaller than the Hubble scale, the effects of dynamical gravity are negligible. Let's therefore consider the case in which matter is minimally coupled to the  metric
\be \label{expWeyl}
g_{\mu\nu} = e^{2 \pi} \eta_{\mu\nu} \; ,
\ee
where the exponential is just a convenient non-linear completion of eq.~(\ref{Weyl1}), and we ``froze'' dynamical gravity, $\hat g_{\mu\nu}= \eta_{\mu\nu}$.
Then $\pi$ non-linearly realizes the conformal group:
\begin{itemize}

\item
Dilations
\be \label{dilation}
D: \qquad \pi(x) \to \pi'(x) \equiv \pi (  \lambda  x ) + \log \lambda \; ;
\ee

\item
Infinitesimal special conformal transformations
\be \label{special}
K_\mu  : \qquad  \pi(x) \to \pi'(x) \equiv \pi \big( x + (c \, x^2 - 2 (c \cdot x) x) 
\big) - 2 c_\mu x^\mu \; ;
\ee

\item
Translations
\be
P_\mu : \qquad \pi(x) \to  \pi'(x) \equiv \pi (x+a) \; .
\ee

\item
Boosts
\be
M_{\mu\nu} : \qquad \pi(x) \to  \pi'(x) \equiv \pi ( \Lambda \cdot x) \; .
\ee

\end{itemize}

The three possible maximally symmetric solutions (AdS, Minkowski, dS) correspond to different unbroken combinations of the above generators.
Clearly,  Minkowski space corresponds to a trivial (constant) configuration for $\pi$, for which translations and boosts are unbroken, whereas dilations and special conformal transformations are spontaneously broken. More interesting are the other solutions. The deSitter solution is  \cite{fubini}
\be \label{dS}
e^ {\pi_{\rm dS}} = \frac{1}{1+ \sfrac14 H^2 \, x^2} \; , 
\ee
which manifestly leaves boosts unbroken. It is also invariant under the combined (infinitesimal) action of $K_\mu$ and $P_\mu$,
\be \label{Smu}
S_\mu = P_\mu -\sfrac14 H^2 K_\mu: \qquad  \pi(x) \to \pi'(x) \equiv \pi \big( x + a -\sfrac14 H^2(a \, x^2 - 2 (a \cdot x) x)
\big) + \sfrac12 H^2 \, a_\mu x^\mu \; ,
\ee
as can be straightforwardly checked. Indeed the four generators $S_\mu$ together with the boost generators $M_{\mu\nu}$ form the full algebra of $SO(4,1)$, the deSitter symmetry group.
Likewise, the AdS configuration corresponds to
\be
e^ {\pi_{\rm AdS}} = \frac{1}{1- \sfrac14 k^2 \, x^2} \; , 
\ee
which preserves boosts as well as the combination $R_\mu = P_\mu + \sfrac14 k^2 K_\mu$. The unbroken symmetries make up the AdS symmetry group, $SO(3,2)$
\cite{fubini}.

If we start from a conformally invariant Lagrangian for $\pi$, we may generically expect one of these configurations to describe the vacuum.
Which one will be a solution to the dynamics, and with what curvature scale, clearly depends on the Lagrangian.
However, we just have to ask whether we can generate all the Galilean-invariant terms we need for a locally deSitter solution to exist, be stable, and admit stable spherically symmetric perturbations, starting from a conformally invariant Lagrangian. If we can, then by symmetry the deSitter solution will exist globally, and will have everywhere the same local properties (stability, etc.).

To construct conformally invariant Lagrangians, we simply write down all possible diff-invariant Lagrangian terms involving the fictitious metric (\ref{expWeyl}). In fact, the conformal group is a subgroup of diffeomorphisms---the subgroup that leaves the metric conformally flat. The simplest term is obviously
\be
\int \! d^4 x \sqrt{-g} = \int \! d^4 x \, e^{4 \pi} \; ,
\ee
which corresponds, upon linearization in $\pi$ to the term proportional to $\pi$ in the galileon lagrangian of the previous sections.  It is clearly conformal invariant, upon changing the integration variable. For instance, under special conformal transformations, eq.~(\ref{special}), we get
\be
\int \! d^4 x \, e^{4 \pi (x)} \to \int \! d^4 x \, e^{4 \pi (x+( c \, x^2 -2(c \cdot x) x) )} \, e^{-8 c_\mu x^\mu} = \int \! d^4 x' \, e^{4 \pi (x')} \; .
\ee 
The next terms, in a derivative expansion, are the curvature invariants $R$, $R^2$, $R_{\mu\nu}R^{\mu\nu}$, $R^\mu {}_{\nu\rho\sigma} R_\mu {}^{\nu\rho\sigma}$, and so on. Let's look more closely at their structure in terms of $\pi$. The Riemann tensor for the conformally flat metric eq.~(\ref{expWeyl}) is schematically
\be
R^\mu {}_{\nu\rho\sigma} \sim \di \pi \di \pi  + \di^2 \pi \; .
\ee
Thus a generic $n$-th order curvature invariant is of the form
\bea
\int \! d^4 x \sqrt{-g} R^n & \sim&  \int \! d^4 x \, e^{(4-2n) \pi} (\di \pi \di \pi  + \di^2 \pi)^n \nonumber \\
& \sim &  \int \! d^4 x \, e^{(4-2n) \pi} \big[ (\di \pi \di \pi)^n + \dots + \di \pi \di \pi \,(\di^2 \pi)^{n-1} +  (\di^2 \pi)^n \big] \label{Rtothen}
\eea
The Galilean invariants we constructed in the previous sections correspond to the second-last term in eq.~(\ref{Rtothen}). The terms with more $\pi$'s and fewer derivatives per $\pi$---those on the left of the Galilean one in eq.~(\ref{Rtothen})---are less important at short distances and small values of $\pi$.
For instance on the deSitter configuration $\pi \simeq -\frac14 H^2 x^2$, the relative importance of one such term with respect to the Galilean one is
\be
\frac{(\di \pi \di \pi)^m \,(\di^2 \pi)^{n-m}}{\di \pi \di \pi \,(\di^2 \pi)^{n-1}} \sim (H^2 \, x^2)^{m-1} \; ,
\ee
which is negligible at sub-horizon scales. The same conclusion obviously holds for the exponential prefactor in eq.~(\ref{Rtothen}), which at lowest order we can set to one.
In fact  at short distances and at small $\pi$-values, conformal invariance reduces to Galilean invariance. This is manifest from expanding dilations (eq.~(\ref{dilation})) and special conformal transformations (eq.~(\ref{special})) at lowest order in $\pi$ and $x$.

However by the same token, at short distances and for small $\pi$ the Galilean term in eq.~(\ref{Rtothen}) is negligible with respect to the one on its right, which has one less $\pi$ and two derivatives acting on every $\pi$. For instance their ratio on the deSitter solution is of order $H^2 x^2$. The terms with two derivatives per $\pi$ are exactly the terms we did not want in sect.~\ref{structure}, because they lead to fourth-order equations of motion, and thus to ghosts.
Therefore, we need to find combinations of curvature invariants for which these unwanted higher-derivative contributions vanish. For such combinations, the most important Lagrangian terms at short distances will be the Galilean ones, and our local analysis of the previous sections will apply.
Since the tuning that sets to zero the $(\di ^2 \pi)^n$ terms corresponds to removing a degree of freedom, terms of this type (by definition of perturbation theory)
will consistently be generated at the quantum level only with a coefficient 
with a small enough size that the ghost will be at the cut-off. This issue is discussed in ref.~\cite{NR}.

The first curvature invariant is
\be \label{R}
\int \! d^4 x \sqrt{-g} R = \int \! d^4 x \, e^{2 \pi} \left[ -6(\di \pi)^2 - 6 \, \Box \pi \right] \; .
\ee
Notice that here we do not need (and we could not either) cancel the $\Box \pi$ term. Indeed  integrating by parts
\be
e^{2 \pi} \, \Box \pi \to - 2 \,  e^{2 \pi} (\di \pi)^2 \; ,
\ee
and so eq.~(\ref{R}) is nothing but the $\pi$ kinetic term.

The next invariant is
\be \label{R2}
\int \! d^4 x \sqrt{-g} \left[ a\, R^2 + b \, R_{\mu\nu} R^{\mu\nu} + c \, R^\mu {}_{\nu\rho\sigma} R_\mu {}^{\nu\rho\sigma} \right]\; .
\ee
Here we have a complication. Upon integration by parts, the three terms above yield all the  same structure in terms of $\pi$
\be \label{R2combination}
\sqrt{-g} \times R^2, (R_{\mu\nu})^2, (R^\mu {}_{\nu\rho\sigma} )^2 \quad \propto
\quad \big[  (\di \pi)^4 + 2 (\di \pi)^2 \, \Box\pi + (\Box \pi)^2
 \big] \; ,
\ee
where $(\di \pi)^4 \equiv ((\di \pi)^2)^2$.
Hence, there is no way of tuning the coefficients in eq.~(\ref{R2}) as to make the $(\Box \pi)^2$ term vanish while keeping a non-trivial Galilean term. However this is an accident peculiar to 4D. We can ask whether there is a suitable combination of invariants in $d$ dimensions, for which the $(\Box \pi)^2$ vanishes and which is regular and non-trivial in the $d \to 4$ limit---a sort of derivative with respect to $d$. In $d$ dimensions the Ricci tensor and scalar read
\bea
R_{\mu\nu} & = & (d-2) \di_\mu \pi\di_\nu \pi - (d-2)  \eta_{\mu\nu} (\di \pi)^2
 - (d-2) \di_\mu \di_\nu \pi - \eta_{\mu\nu} \Box \pi \\
R & = & - e^{-2\pi} \left[ (d-1)(d-2) (\di \pi)^2 + 2(d-1) \Box \pi \right]
\eea
(we will not need the Riemann tensor). Then for instance the combination
\bea \label{d-dim}
& & \int \! d^d x \sqrt{-g_d} \, \frac{1}{d-4} \left[ \frac{R_{\mu\nu} R^{\mu\nu}}{d-1} - \frac{R^2}{(d-1)^2} \right] =\nonumber  \\
&=&  \int \! d^d x \, e^{(d-4) \pi} \left[ \frac{(d-2)(3d-4)}{2(d-1)}(\di \pi)^2 \Box \pi + \frac{(d-2)^3}{2(d-1)} (\partial \pi)^4 + (\Box \pi)^2 \right] \; .
\eea
is regular for $d \to 4$, and linearly independent of eq.~(\ref{R2combination}).
By combining eqs.~(\ref{R2combination}) and (\ref{d-dim}) we can tune to zero the $(\Box \pi)^2$ while retaining the Galilean interaction (as well as the subleading $(\di \pi)^4$ term).

Alternatively, we can shelve for a second the auxiliary metric $g_{\mu\nu} = e^{2 \pi} \eta _{\mu\nu}$, and look for invariants in 4D directly in terms of $\pi$. At the four-derivative level (which corresponds to $R^2$ etc.), invariance under dilations (eq.~(\ref{dilation})) tells us that there is no $e^{n \pi}$ overall factor. Then, under special conformal transformations (eq.~(\ref{special})) the individual terms transform as
\bea
{\textstyle \int \! d^4 x} \, (\di \pi)^4 & \to & {\textstyle \int \! d^4 x} \, \big [ (\di \pi)^4 - 8 c^\mu \di_\mu \pi (\di \pi)^2 \big] \\
{\textstyle \int \! d^4 x} \, (\di \pi)^2 \Box \pi & \to & {\textstyle \int \! d^4 x} \, \big[ (\di \pi)^2 \Box \pi + 4 c^\mu \di_\mu \pi (\di \pi)^2 - 4 c^\mu \di_\mu \pi \Box \pi \big]\\
{\textstyle \int \! d^4 x} \, (\Box \pi)^2 & \to & {\textstyle \int \! d^4 x} \, \big[ (\Box \pi)^2  +8 c^\mu \di_\mu \pi \Box \pi \big]
\eea
where on the r.h.s.'s we changed integration variable, $x^\mu \to x^\mu + (c^\mu x^2 - 2 (c \cdot x) x^\mu )$.
We thus see that at the Lagrangian level the only invariant combination is precisely eq.~(\ref{R2combination}). However, if we require that the e.o.m.~be invariant, the Lagrangian need only be invariant up to a total derivative. Then the $(\Box \pi)^2$ term is invariant by itself, since $\di_\mu \pi \Box \pi$ is a total derivative---see e.g.~eq.~(\ref{deltaDGP}). This means that we can set it to zero while preserving conformal invariance. 

Notice in passing that Lagrangian terms of the form $e^{(4 - 2n) \pi} (\di^2 \pi)^m (\di \pi)^{2n-2m}$ are automatically invariant under dilations, eq.~(\ref{dilation}), whereas as we saw only some very specific combinations of them are invariant under special conformal transformations as well. This goes against the common expectation that scale invariance alone in enough to guarantee full conformal invariance. In fact, the standard arguments do not apply for higher-derivative terms---see e.g.~ref.~\cite{polchinski}.

At the next orders,  $R^3$ and $R^4$, there are no complications. Suitable combinations of curvature invariants involving just the Ricci tensor and scalar give the Galilean terms we want, while tuning the unwanted $(\Box \pi)^3$ and $(\Box \pi)^4$ terms to zero. At order $R^3$ we get (after integration by parts)
\be
 \sqrt{-g} \left[ \sfrac{7}{36} \, R^3 +  R (R_{\mu\nu})^2
- (R_{\mu\nu})^3 
\right] =   e^{-2\pi} \big[ 24 \, {\cal L}_4 + \dots  \big] \; ,
\ee
where ${\cal L}_4$ is the fourth-order Galilean invariant, eq.~(\ref{Galileo4}), and the dots stand for terms with fewer derivative per $\pi$, which are negligible at short distances. Likewise, at order $R^4$
\be
\sqrt{-g} \left[ \sfrac{93}{2 \cdot 6^4} \, R^4 - \sfrac{39}{4 \cdot 6^2} \, R^2 (R_{\mu\nu})^2 + \sfrac5{12}  \, R (R_{\mu\nu})^3 + \sfrac3{16} \big((R_{\mu\nu})^2\big)^2 -\sfrac{3}{8} (R_{\mu\nu})^4 
\right] =  e^{-4\pi} \big[ 16 \, {\cal L}_5 + \dots  \big] \; .
\ee

In conclusion, we can reproduce any Galilean-invariant Lagrangian of the form
eq.~(\ref{Lsum}) as the short-distance limit of a globally defined, conformally invariant Lagrangian for $\pi$. Then, the existence of a deSitter solution as well its properties (stability, etc.) can be studied locally, about any point, as we did in the previous section. The symmetries of the full Lagrangian guarantee that the deSitter solution we find locally can be promoted to a global one, and the isometries of deSitter space guarantee that the stability and the other local properties of the solution will be the same throughout space.

\subsection{Gravitational back-reaction}\label{gravity}

We now consider the (phenomenologically interesting) case where gravity is dynamical, $M_{\rm Pl} < \infty$. In our notation, this amounts to changing the metric seen by matter, eq.~(\ref{expWeyl}) into
\be \label{full_metric}
g_{\mu\nu} = e^{2 \pi}  \hat g_{\mu\nu} \; ,
\ee 
of which eq.~(\ref{Weyl1}) is the small field limit, and making $\hat g_{\mu\nu}$ dynamical. Also we should covariantize the $\pi$ action, which corresponds to coupling $\pi$ to $\hat g_{\mu\nu}$. As usual there are many inequivalent ways to do so, parameterized by non-minimal couplings of $\pi$ to curvature invariants. Now, if we want our deSitter solution to survive after the coupling to gravity is taken into account, we have to require that gravitational interactions do not spoil the conformal invariance of the $\pi$ action---so that the existence of a deSitter solution is guaranteed by symmetry. 
But the only way to couple a conformal theory to gravity while preserving conformal invariance is imposing that the action be invariant under Weyl rescalings of the metric:
\be
\hat g_{\mu\nu} (x) \to \hat g'_{\mu\nu}(x) = e^{2 \omega(x)} \, \hat g_{\mu\nu}(x) \; .
\ee
If we do so, the dilaton of the CFT and the metric have to appear precisely in the combination (\ref{full_metric}), so that a conformal transformation on $\pi$ may be reabsorbed into a Weyl transformation of $\hat g_{\mu\nu}$.
This corresponds to gauging the conformal group. In fact, this is how we constructed the conformally-invariant action for  $\pi$ in the first place---by making the conformal group that acts on $\pi$ a subgroup of diffeomorphisms acting on $g_{\mu\nu}$. The full action is then the sum of the manifestly covariant operators of last section, where now $R^\mu {}_{\nu\rho\sigma}$ stands for the Riemann tensor of the metric (\ref{full_metric}).\footnote{The combination in eq.~(\ref{d-dim}) is non-singular in $d=4$ only for conformally flat metrics, as manifest from its l.h.s. This means that the DGP-like term 
\be
2(\di \pi)^2 \Box \pi + (\di \pi)^4
\ee
cannot be coupled to gravity in a Weyl invariant fashion. Given that---as we will see shortly---the Weyl invariant couplings we are constructing are not viable, we do not investigate this issue further. Also notice that when using a generic metric $g_{\mu\nu}$ rather than the conformally flat one of last section, we have new independent higher-order curvature invariants, involving for instance the Riemann tensor---which for conformally flat metrics is instead determined by the Ricci tensor.
}
As a result, our dilaton  $\pi$ disappears from the action---nowhere do $\pi$ and $\hat g_{\mu\nu}$ appear separately from each other. In fact, via a Weyl transformation we can choose `unitary gauge', 
\be
\pi = 0 \; , \qquad g_{\mu\nu} = \hat g_{\mu\nu} \; .
\ee
The action now is just a higher-derivative action for the metric, coupled to matter.
The  problem with this action, is that the energy scale that ``suppresses'' the higher-derivative terms is minuscule. Indeed, in the presence of an Einstein-Hilbert term $\mpl^2 R$, in order to have a deSitter solution with curvature $\sim H_0^2$, we need either a non-derivative operator
\be
\sqrt{-g} \, \Lambda \sim \sqrt{-g} \, \mpl^2 H_0^2 \; ,
\ee
which is not surprising news, or higher-order curvature invariants weighed by inverse powers of $H_0^2$,
\be \label{higher_R}
\sqrt{-g} \, \frac{\mpl^2}{H_0^{2n-2}} R^n \; .
\ee
The latter possibility, which is the non-conventional one, corresponds to a {\em ultraviolet} modification of GR at curvatures of order $H_0^2$, which, even at the classical level, is clearly ruled out by experiments.

We are thus led to the conclusion that viable gravitational interactions of $\pi$ cannot preserve conformal invariance---a conclusion that is also strengthened by Weyl invariance's being generically anomalous at the quantum level.  As a consequence, the deSitter solution we found for $\pi$ in general will be spoiled when the gravitational backreaction is taken into account. 
Also notice that the problem of having higher-derivative terms for the metric with huge coefficients like in eq.~(\ref{higher_R}), will be shared by a large class of Lagrangians where $\pi$ has important non-minimal couplings to $\hat g_{\mu\nu}$. The safest possibility from this viewpoint seems to be minimal coupling---it involves the least number of derivatives of $\hat g_{\mu\nu}$ compatible with diff-invariance.
Let's therefore consider the case of minimal coupling. This amounts to covariantize the $\pi$ action just by contracting Lorentz indices with $\hat g^{\mu\nu}$, replacing ordinary derivatives with covariant ones, and integrating in $d^4 x \sqrt{-\hat g}$. Also, we include an Einstein term $\mpl^2 \hat R$, to reproduce Newton's law. Scaling the various terms in the galilean limit like in eq.~(\ref{pisquiggly}) and using the results of section 3.1 for their relation to the other terms,
the $n$-th order Lagrangian term is schematically
\be \label{Lagrangian_f}
{\cal L} \sim \frac{f^2}{H_0^{2n-2}} e^{(4-2 n) \pi} (\di \pi \di \pi + \di \di \pi)^n \; .
\ee

Now, as we saw, a deSitter configuration for $\pi$ corresponds to eq.~(\ref{dS}). For cosmology, it is more convenient to use coordinates where the solution only depends on (conformal) time,
\be \label{dS_cosmo}
e^{\pi _{\rm dS}} = -\frac{1}{H_0 \eta} \; .
\ee
The relation between this and eq.~(\ref{dS}) is just a special conformal transformation on $\pi$ combined with a translation. However, as we discussed, when gravity is taken into account conformal transformations are no longer a symmetry. Therefore the two expressions for the deSitter configuration are physically inequivalent---i.e.~they produce different gravitational fields. 
If we are looking for FRW solutions for the metric, we have to stick to eq.~(\ref{dS_cosmo}).
The stress-energy tensor scales like the Lagrangian (\ref{Lagrangian_f}), which for our deSitter configuration gives
\be
T_{\mu\nu} \sim \frac{1}{\eta^4} \frac{f^2}{H_0^2} \; .
\ee
The curvature perceived by matter is a combination of that due to its conformal coupling to $\pi$, and that of $\hat g_{\mu\nu}$, which in turn is sourced by the $\pi$ stress energy tensor:
\bea
R_{\mu\nu} & \sim & \di \pi \di \pi + \di \di \pi + \hat R_{\mu\nu}   \\
		& \sim &		\frac{1}{\eta^2} \left[ 1 + \frac{f^2}{\mpl^2} \frac{1}{H_0^2 \eta^2} \right] \; .
\eea
The latter contribution becomes sizable and spoils the deSitter geometry at conformal times
\be
|\eta| \lesssim H_0^{-1} \cdot \frac{f}{\mpl} \; .
\ee
If we could make $f$ smaller and smaller, we could postpone this moment at will. However we have already shown that tests of GR at subcosmological scales constrain $f$ to be not significantly smaller than $M_P$. So we cannot play with $f$ to make the solutions more stable. 
Anyway, since proper time $t$ depends only logarithmically on $\eta$, this means that the geometry will look approximately deSitter only for times of order Hubble. That is, gravitational interactions necessarily spoil the deSitter solution at Hubble scales.

\section{Discussion and Outlook}\label{discussion}

We conclude with a few remarks on our results and on possible future developments. 
First, let's summarize our findings. We have considered theories that modify the infrared behavior of gravity where, at least in a local patch smaller than the cosmological horizon:
\begin{itemize}
\item[{\em (i)}] The modification of gravitational interactions is due to an effective relativistic scalar degree of freedom $\pi$ kinetically mixed with the metric, or equivalently, universally coupled to matter via the interaction $\pi \, T^\mu {}_\mu$.
\footnote{
This is the case for instance in Fierz-Pauli massive gravity and in the DGP model, whereas massive gravity theories that crucially rely on spontaneous breaking of Lorentz invariance are left out.
}
\item[{\em (ii)}] This scalar degree of freedom generically decouples from matter at shorter scales thanks to important derivative self-interactions (Vainshtein effect).
\footnote{
Some form of non-linearities are necessary for screening a universally coupled scalar at short scales, thus recovering agreement with solar system tests of GR. The only robust alternative to the Vainshtein effect we are aware of is the Chameleon mechanism \cite{chameleon}. There, the non-linearities are in the coupling of the scalar to matter, rather than in its self-interactions.}
\end{itemize}
In this class of theories, we have argued that the dynamics of $\pi$ must enjoy an internal Galilean symmetry.  In 4D there are only five possible Lagrangian terms with such a symmetry. This makes a thorough analysis of very symmetric solutions possible. First, we demanded the existence and stability of a deSitter solution in the absence of a cosmological constant---a self-accelerating solution {\em \`a la} DGP, but with no ghost instabilities.
Then, we imposed that with these deSitter asymptotics there exist spherically symmetric Vainshtein-like solutions, describing the $\pi$ configuration about compact sources, 
and that these configurations are also stable against small perturbations.  We found some tension, if no contradiction, in satisfying these requirements. It is plausible---although we have no proof of this---that demanding the existence and stability of more complicated solutions, like e.g.~with several compact sources in generic positions,
may lead to contradictions.
However, we also found that small excitations are {\em forced} to be superluminal at {\em large} distances from a compact source, where $\pi$ is in the linear regime, even if we don't require the existence of self-accelerating solutions.  


We ran into trouble also as a result of the opposite problem: the very slow speed of angular fluctuations when approaching the source due to the presence of the galilean terms $\mathcal{L}_{4,5}$. This makes  a static computation of the background not trustable because of huge retardation effects, which are also at the origin of the very low energy cutoff of the theory.
To avoid these difficulties we considered the subclass of Lagrangians with $c_4=c_5=0$. This requirement is still compatible with the absence of ghosts around deSitter background and with the existence a stable Vainshtein-like solution around compact sources. The kinetic term in the angular directions is now enhanced, as a result quantum fluctuations are suppressed and the effective cutoff increases like in DGP above the naive estimate $\Lambda\sim (1000 \: \rm{km})^{-1}$.    


Finally, we considered whether our local galileon $\pi$ can be promoted to a global scalar degree of freedom, coupled someway to four-dimensional GR. First neglecting gravity, we promoted the Poincar\'e group and our local galilean invariance to the conformal group, with $\pi$ as the dilaton. The deSitter group is one of the maximal subgroups of the conformal group, and the existence of deSitter-like solutions for $\pi$ may be generically expected by symmetry considerations. We showed that our local Galilean invariant terms are in fact the short-distance limit of globally defined, conformally-invariant Lagrangian terms for $\pi$. The problem however is in coupling this system to gravity. This necessarily breaks conformal invariance, and as a consequence spoils the deSitter geometry at times of order of the Hubble time, and longer. Although this is not a disaster---in fact this model may well yield a realistic cosmology, for we have no idea of whether our universe's future resembles deSitter space---the symmetry motivation for the `conformal completion' of the Galilean symmetry is somewhat lost.

Let us stress once again that our local analysis of the Galilean model is quite generic---it applies to all theories obeying conditions {\em (i)} and {\em (ii)} above.
Therefore, the partial failure of our attempt to `IR-complete' the Galilean model should not be seen as impairing the local analysis.
Our local analysis applies for instance to the DGP model---as we showed in sect.~\ref{DGPcosmo}. DGP at short distances corresponds to a Galilean model with non-vanishing  $c_2$ and $c_3$ only.
There however, it is manifest that the IR completion is not a four-dimensional local theory---it is in fact a five-dimensional one, and our $\pi$ degree of freedom stops making sense at scales of order of the cosmological horizon, where its interactions and mixings with the whole continuum of KK modes cannot be neglected anymore.
We do not know whether this is going to be the case for a generic Galilean model, or whether instead there exist purely four-dimensional local IR completions that realize the deSitter symmetry group as a subgroup of a larger internal symmetry group, other than the conformal one. Perhaps, as we suggest in sect.~\ref{symmetry}, there exists one such theory where the internal symmetry group is the 5D Poincar\'e group. This in a sense would yield a purely four-dimensional version of DGP, without the extra complications of 5D gravity.

On the other hand, since DGP locally reduces to a particular Galilean model, perhaps a generic one can be IR completed by a generalization of DGP.  Possible generalizations include higher-than-five dimensional setups\cite{deRham:2007xp}, or the standard five-dimensional setup with non-minimal bulk or brane operators. In the latter case, we might try adding higher-curvature invariants in the bulk, but in order to keep second-order e.o.m.~for $\pi$---a crucial feature of our galilean-invariant Lagrangians---we are only allowed to use Lovelock combinations. But in 5D we just have one such combination, whereas to reproduce a generic Galilean model we need several independent terms. The other possibility  is to add localized terms on the brane involving powers of the extrinsic curvature. Indeed for small gravitational fields the extrinsic curvature is $K_{\mu\nu} \propto \di_\mu\di_\nu \pi$, so such terms will directly yield derivative self-interactions for $\pi$. The problem with this is that they involve too many derivatives---two per $\pi$ field. Another possibility would be to add new five-dimensional degrees of freedom besides gravity, or, as we said, to consider higher-dimensional spacetimes. We leave this for future investigations.

We now turn our attention to the issue of the {\em ultraviolet} completion. From this viewpoint, our model is just a generalization of the 4D effective action of DGP that gets rid of the ghost on the self-accelerated branch, but that nevertheless shares with DGP the other pathologies identified in ref.~\cite{adams}. Namely, {\em (a)} the  superluminalilty of scalar excitations about some physically relevant solutions, and {\em (b)} the excessive softness of certain scattering amplitudes at low energies.
These features are in fact related to each other, and although they do not necessarily signal any inconsistency of the low-energy effective theory---unless closed time-like curves may form---they do indicate that our model cannot arise as the low-energy description of a Lorentz-invariant microscopic theory obeying the standard axioms of $S$-matrix theory, like e.g.~any renormalizable relativistic QFT \cite{adams}. This is to be contrasted with what happens for GR. The Weinberg-Witten theorem  makes the quest for a UV-completion of GR within local Lorentz-invariant QFTs particularly hard---perhaps hopeless \cite{WW}. Nevertheless, looking at GR as a perturbative Lorentz-invariant effective theory of gravitons living in Minkowski space, one finds that gravitational interactions cannot lead to superluminal propagation with respect to the underlying Minkowski light-cone as long as sources obey the null energy condition \cite{VBL, adams}.
Therefore, although probably not local, the UV completion of GR may well be Lorentz-invariant (like e.g.~string theory).

From a more theoretical perspective, we have constructed a class of models with peculiar derivative interactions. On the one hand, $\pi$'s self-interactions involve many derivatives, more than one derivative per field. On the other hand, the particular combinations we isolated yield second-order equations of motion. At the classical level, this means that  no new ghost-like degree of freedom shows up in the non-linear regime. This is a necessary condition 
for the quantum effective theory not to break down for sizable classical non-linearities. In fact with a consistent, if not generic choice of counter-terms, our effective theories can be extrapolated deep into the non-linear regime \cite{NR}. In a forthcoming publication, we show that these peculiarities allow our models to violate the null energy condition with no manifest pathologies in the low-energy theory. In particular, our models subtly evade the theorems of refs.~\cite{Dubovsky:2005xd, luty}.


\section*{Acknowledgements}
We would like to thank G. Dvali and  L. Hui for stimulating discussions.
A.N.~and E.T.~would like to thank the Institute of Theoretical Physics of the Ecole Polytechnique F\'ed\'erale de Lausanne for hospitality during this project. The work of R.R. is partially supported by the Swiss National Science Foundation under contract No. 200021-116372.


\appendix
\section*{Appendix}

\section{Existence and uniqueness of galilean-invariant terms at each order in $\pi$ }

As we discussed in section \ref{structure}, the generic structure of a galilean-invariant lagrangian term with $n+1$ $\pi$-factors is $(\partial^2 \pi)^{n-1} \partial \pi \partial \pi$ and the corresponding equation of motion has exactly two derivatives acting on every $\pi$, $(\partial \partial \pi)^n =0$. However, because of the shift symmetry of the $\pi$ Lagrangian, the e.o.m. for $\pi$ can also be written as the divergence of the conserved current associated to this symmetry, schematically: $\partial^\mu (\partial^{2n-1} \pi^n)_\mu =  (\partial \partial \pi)^n =0$. This equivalence can be read as the statement that the combination $(\partial \partial \pi)^n$ that appears in the e.o.m. derived from a galilean-invariant term is a total derivative.

In the following we show that there exists one and only one total derivative $(\partial \partial \pi)^n$ for every $n$; for this reason there can be at most one galilean-invariant term at each order and eventually we conclude that such a term exists for every $n$ by constructing it explicitly, starting from the corresponding total derivative. 

First of all, notice that it's easy to write down a total derivative at the $n$-th order:  
\be\label{totder}
\sum_p (-1)^p \eta^{\mu_1 p(\nu_1)} \ldots \eta^{\mu_n p(\nu_n)} 
\: \di_{\mu_1} \di_{\nu_1} \pi \ldots \di_{\mu_n} \di_{\nu_n} \pi \equiv  T^{{\mu_1}{\nu_1} \ldots {\mu_n} {\nu_n}}  \di_{\mu_1} \di_{\nu_1} \pi \ldots \di_{\mu_n} \di_{\nu_n} \pi\, ,
\ee
here the sum runs over all the permutations $p$ of the indices $\nu_i$ and $(-1)^p$ is the parity of the permutation. We now prove that this is the only one for a given $n$. In what follows we use the same notation of section \ref{structure}: $\Pi$ is the matrix of second derivatives of $\pi$, $\Pi^{\mu}_\nu= \partial^\mu \partial_\nu \pi$ and the brackets $[{\cal M}]$ denote the trace of the martrix ${\cal M}$, hence $[\Pi^n]$ is the cyclic contraction of order $n$.

Supppose that there are two total derivatives with the same $n$.  
Take the linear combination ${\cal L}_n$ of the two such that the terms with the highest number $m\leq n-1$ of $\Box \pi$ are $c_m^i (\Box \pi)^m [\Pi^{i_1}] \ldots [\Pi^{i_k}]$ with $i_1 + \ldots+ i_k=n-m$ and $i_j\geq 2$ for $j=1,\ldots,k$: actually we can always cancel in ${\cal L}_n$ at least the term $(\Box \pi)^n$. The multi-index $i$ counts the different contractions that can appear in $(\partial \partial \pi)^{n-m}$.
 
Now if we consider the combination ${\cal L}_n$ as a Lagrangian density, by definition of total derivative the equation of motion for $\pi$ that we get from ${\cal L}_n$ must vanish. Let's see what are the consequences of this condition on the coefficients $c_m^i$. To find the e.o.m. we can integrate by parts twice and isolate a $\pi$ shifting the two derivatives acting on it to all the other $\Pi$ in turn, in particular when we choose a $\pi$ in the $j$-th cyclic contraction and we let the two derivatives act on the $\pi$ close to the one we are starting with in the same cycle, we end up with the structure $[\Box \Pi \ldots \Pi^{i_{j}-2}]$.
Therefore among many terms, ${\cal L}_n$ will give a contribution to the e.o.m. proportional to 
\be\label{eomtd}
c_m^i (\Box \pi)^m [\Pi^{i_1}] \ldots [\Pi^{i_{j-1}}] [\Box \Pi \ldots \Pi^{i_{j}-2}] [\Pi^{i_{j+1}}] \ldots [\Pi^{i_k}].
\ee
In order to have a vanishing equation of motion, this must be cancelled by a contribution coming from another term in ${\cal L}_n$. The only possible contraction that gives something proportional to (\ref{eomtd}) in the equation of motion would be given by
\be\label{altrotermine}
{\cal L}_n \supset (\Box \pi)^{m+1} [\Pi^{i_1}] \ldots [\Pi^{i_{j-1}}] [\Pi^{i_{j}-1}] [\Pi^{i_{j+1}}] \ldots [\Pi^{i_k}].
\ee
In facy if we integrate by parts and we shift the $\Box$ of the $(m+1)$-th $\Box \pi$ to one of the $\Pi$ in the $j$-th cycle we get exactly $\pi \ldots$(\ref{eomtd}): this is the only other possibility to have in the e.o.m. a $\Box$ in the structure $[\Box \Pi \ldots \Pi^{i_{j}-2}]$, maintaining at the same time $(\Box \pi)^{m}$ and the cycles $i_1,\ldots,i_{j-1},i_{j+1},\ldots, i_{k}$.

However, terms with $(\Box \pi)^{m+1}$ like (\ref{altrotermine}) are absent from ${\cal L}_n$ by construction, and as a consequence ${\cal L}_n$ can be a total derivative only if $c_m^i=0$. By induction on $m$ we conclude that the whole ${\cal L}_n$ must be zero: the two total derivatives we started with are the same a part from an irrelevant overall constant factor. In other words, the structure $(\partial \partial \pi)^n$ can be a total derivative only if it contains $\Box \pi$ to {\it all} powers from $0$ to $n$, and this can happen only once for a given $n$. 

Since we proved that the only total derivative for every $n$ is given by (\ref{totder}), the last step of our argument is to show that this specific contraction can be always derived as the equation of motion of a galilean-invariant term. The galilean term can be constructed by replacing in turn each derivative $\partial_{\mu_i}$ with $\partial_{\mu_i} \pi$:
\be\label{galinv}
 {\cal L}_{\rm{galilean}}= T^{{\mu_1}{\nu_1} \ldots {\mu_n} {\nu_n}}  (\di_{\mu_1} \pi \, \di_{\nu_1} \pi \, \di_{\mu_2}  \di_{\nu_2} \pi\ldots \di_{\mu_n} \di_{\nu_n} \pi + \ldots + \di_{\mu_1}  \di_{\nu_1} \pi \ldots \di_{\mu_{n-1}} \di_{\nu_{n-1}} \pi \,  \di_{\mu_n} \pi \,  \di_{\nu_n} \pi   )\, .
\ee
It's easy to check that the equation of motion we find starting from (\ref{galinv}) is precisely $-2n  \, T^{{\mu_1}{\nu_1} \ldots {\mu_n} {\nu_n}}$ $ \di_{\mu_1} \di_{\nu_1} \pi \ldots \di_{\mu_n} \di_{\nu_n} \pi = 0$; in fact terms with 3 or 4 derivatives acting on the same $\pi$ must have 2 derivatives with a 
$\nu$ index like $\partial_{\nu_{i}}  \partial_{\nu_{j}} \partial_{\mu_k} \pi $ or  $\partial_{\nu_{i}}  \partial_{\nu_{j}} \partial_{\mu_k} \partial_{\mu_l} \pi $ and because of the antisymmetry of $T^{{\mu_1}{\nu_1} \ldots {\mu_n} {\nu_n}} $ in the exchange of $\nu_i \leftrightarrow \nu_j$ due to the factor $(-1)^p$, they are all zero.


\section{$c_i \leftrightarrow d_i$}\label{cprime}

We want to expand the equation of motion about the deSitter solution,
\be \label{expand}
\pi \to -\sfrac14 H^2 \, x_\mu x^\mu + \pi \; , \qquad 
\di_\mu \di_\nu \pi \to -\sfrac12 H^2 \eta_{\mu\nu} +   \di_\mu \di_\nu \pi \; .
\ee
Each Galilean term (\ref{eom1}--\ref{eom5}) will then yield a linear combination of itself and the lower-order ones. The contributions to the tadpole ${\cal E}_1$ coming from all the terms  will cancel---since we are assuming that the deSitter background is a solution to the e.o.m. We will therefore neglect the tadpole and concentrate on the other terms.
Upon the replacement (\ref{expand}) we get
\begin{eqnarray}
{\cal E}_2 & \to & {\cal E}_2  \\
{\cal E}_3 & \to & {\cal E}_3 - 3 H^2 \,{\cal E}_2  \\
{\cal E}_4 & \to & {\cal E}_4 - 3 H^2 \,{\cal E}_3 + \sfrac92 H^4 \,{\cal E}_2   \\
{\cal E}_5 & \to & {\cal E}_5 - 2 H^2 \,{\cal E}_4 + 3 H^4 \,{\cal E}_3 
	- 3 H^6 \,{\cal E}_2
\end{eqnarray}
Therefore, comparing eq.~(\ref{eom}) with eq.~(\ref{eomprimed}) we have
\be
\left[ \begin{array}{c}
d_2 \\ d_3 \\ d_4 \\ d_5 
\end{array} \right] =
\left[  \begin{array}{cccc}
1 & - 3 H^2 & \sfrac92 H^4 & - 3 H^6 \\
0 & 1 & -3 H^2 & 3 H^4 \\
0 & 0 & 1 & -2 H^2 \\
0 & 0 & 0 & 1 
\end{array} \right]
\cdot
\left[ \begin{array}{c}
c_2 \\ c_3 \\ c_4 \\ c_5 
\end{array} \right]
\ee
The inverse relation is
\be
\left[ \begin{array}{c}
c_2 \\ c_3 \\ c_4 \\ c_5 
\end{array} \right] =
\left[  \begin{array}{cccc}
1 & 3 H^2 & \sfrac92 H^4 & 3 H^6 \\
0 & 1 & 3 H^2 & 3 H^4 \\
0 & 0 & 1 & 2 H^2 \\
0 & 0 & 0 & 1 
\end{array} \right]
\cdot
\left[ \begin{array}{c}
d_2 \\ d_3 \\ d_4 \\ d_5 
\end{array} \right]
\ee
Notice that the two matrices that describe the mappings $\{c\} \to \{d\}$ and $\{d\} \to \{c\}$ are exactly the same, upon replacing $H^2$ with $-H^2$. This had to be expected, for we can think of the trivial configuration as an excitation of the deSitter one with $\pi = + \frac14 H^2 x^2 $. Then, like in (\ref{expand}), to get the equation of motion for $\pi$ fluctuations about the trivial configuration we have to perform the replacement
\be
\pi \to +\sfrac14 H^2 x_\mu x^\mu + \pi
\ee
everywhere in the $\pi$ e.o.m.~about the deSitter solution. Of course this way we will not recover the tadpole term, which we can however reconstruct by imposing the existence of a deSitter solution with given $H^2$.


\end{document}